\documentclass{vldb}
\usepackage{graphicx}
\usepackage{balance}  
\usepackage{epsfig}
\usepackage{subfigure}
\usepackage{multirow}
\usepackage{times}
\usepackage{color}
\usepackage{algorithm}
\usepackage[noend]{algorithmic}
\usepackage[mathscr]{eucal}
\usepackage{soul}
\usepackage{enumitem}
\usepackage[hyphens]{url}

\newtheorem{definition}{definition}

\newcommand{\dgcc}{DGCC}
\newcommand{\distdgcc}{DistDGCC}
\newcommand{\aries}{ARIES logging}
\newcommand{\command}{Command logging}

\begin{document}

	\setlength{\pdfpageheight}{\paperheight}
	\setlength{\pdfpagewidth}{\paperwidth}


\title{Scaling Distributed Transaction Processing and Recovery based on Dependency Logging}

	\author{
	    	Chang Yao$^\ddag$,\quad Meihui Zhang$^\S$,\quad  Qian Lin$^\ddag$,\quad  Beng Chin Ooi$^\ddag$,\quad  Jiatao Xu$^\sharp$\\
	    	{$^\ddag$National University of Singapore, $^\S$Singapore University of Technology and Design, $^\sharp$Tencent Inc.}\\
	    	$^\ddag$\{yaochang,linqian,ooibc\}@comp.nus.edu.sg,\enspace $^\S$ meihui\_zhang@sutd.edu.sg \\ \enspace $^\sharp$sunnyxu@tencent.com
    }

	\maketitle

\begin{abstract}	

Dependency Graph based Concurrency Control (DGCC) protocol has been shown to achieve good performance 
on multi-core in-memory system.
\dgcc\ separates contention resolution from the transaction execution 
and employs dependency graphs to derive serializable transaction schedules.
However,
distributed transactions complicate the dependency resolution,  
and therefore, an effective transaction partitioning strategy is essential to reduce expensive multi-node distributed transactions.
During failure recovery,
log must be examined from the last checkpoint onwards and the affected transactions are re-executed 
based on the way they are partitioned and executed.  
Existing approaches treat both transaction management and recovery as two separate problems, 
even though recovery is dependent on the sequence in which transactions are executed.

In this paper, we propose to treat the transaction management and recovery problems as one.
We first propose an efficient \textbf{Dis}tri-\\buted \textbf{D}ependency \textbf{G}raph based \textbf{C}oncurrency \textbf{C}ontrol (\distdgcc) protocol for handling transactions spanning multiple nodes, 
and propose a new novel and efficient logging protocol called \textbf{Dependency Logging} 
that also makes use of dependency graphs for efficient logging and recovery.
\distdgcc\ optimizes the average cost for each distributed transaction
by processing transactions in batch.
Moreover, it also reduces the effects of thread blocking caused by distributed transactions 
and consequently improves the runtime performance.
Further,
dependency logging exploits the same data structure that is used by \distdgcc\ to reduce the logging overhead, 
as well as the logical dependency information to improve the recovery parallelism.
Extensive experiments are conducted to evaluate the performance of our proposed technique against state-of-the-art techniques. 
Experimental results show that \distdgcc\ is efficient and scalable, and
dependency logging supports fast recovery with marginal runtime overhead.
Hence, the overall system performance is significantly improved as a result.
\end{abstract}
	


\section{Introduction}
	
Database systems process transactions~\cite{vldb81:Gray} to effect online updates. 
They serve as the infrastructure of interactive applications such as stock trading, 
banking, e-commerce and inventory management.
Naturally,
OnLine Transaction Processing (OLTP) plays a key role in the database systems as well as applications built on top of it. 
In-memory systems have been gaining tractions in recent years due to factors such as the increased capacity of main memory and its decreased price,  
and the widening gap in memory bandwidth with respect to the disk storage.
Consequently, 
the cost of buffer management is further reduced or even eliminated~\cite{vldb08:Kallman}, 
and the performance of in-memory OLTP systems is now mainly constrained by latching, locking and logging~\cite{sigmod08:harizopoulos, tkde15:zhanghao}.  
Accordingly, many recent research efforts for in-memory OLTP systems have been focusing on the design and optimization of concurrency control protocols~\cite{osdi14:Mu, sigmod16:Ren, sigmod2016:yu} and logging techniques~\cite{icde14:Malviya, sigmod2016:yao, osdi2014:zheng}.

Most of the existing concurrency control protocols resolve contentions by examining the conflicts among individual transactions.
As an optimization, various batching strategies may be employed.
However, batching is traditionally considered to be supplementary in the design of concurrency control protocol.
In contrast, to reduce frequent disk I/Os, 
database systems typically write logs in a batch manner~\cite{sosp1987:robert}.
As a consequence, the tuple-oriented concurrency control and the batched transaction logging are a mismatch that may counter each other's optimizations. 

\begin{figure}[h]
	\subfigure[Traditional two phase commit with logging]{
		\centering
		\label{fig:dist_example_1}
		\includegraphics[scale=0.4]{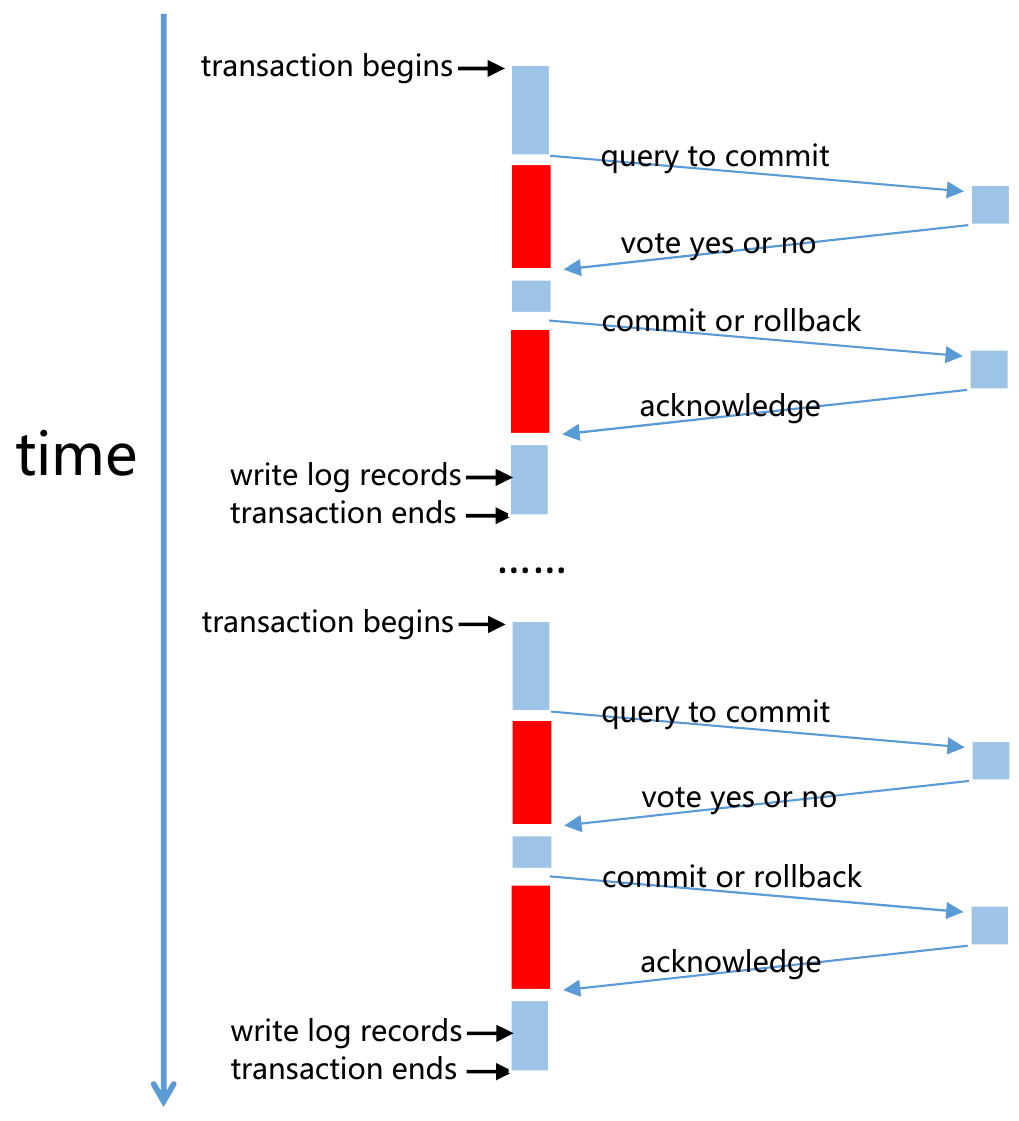}}\quad
	\subfigure[\distdgcc\ with dependency logging]{
		\centering
		\label{fig:dist_example_2}
		\includegraphics[scale=0.4]{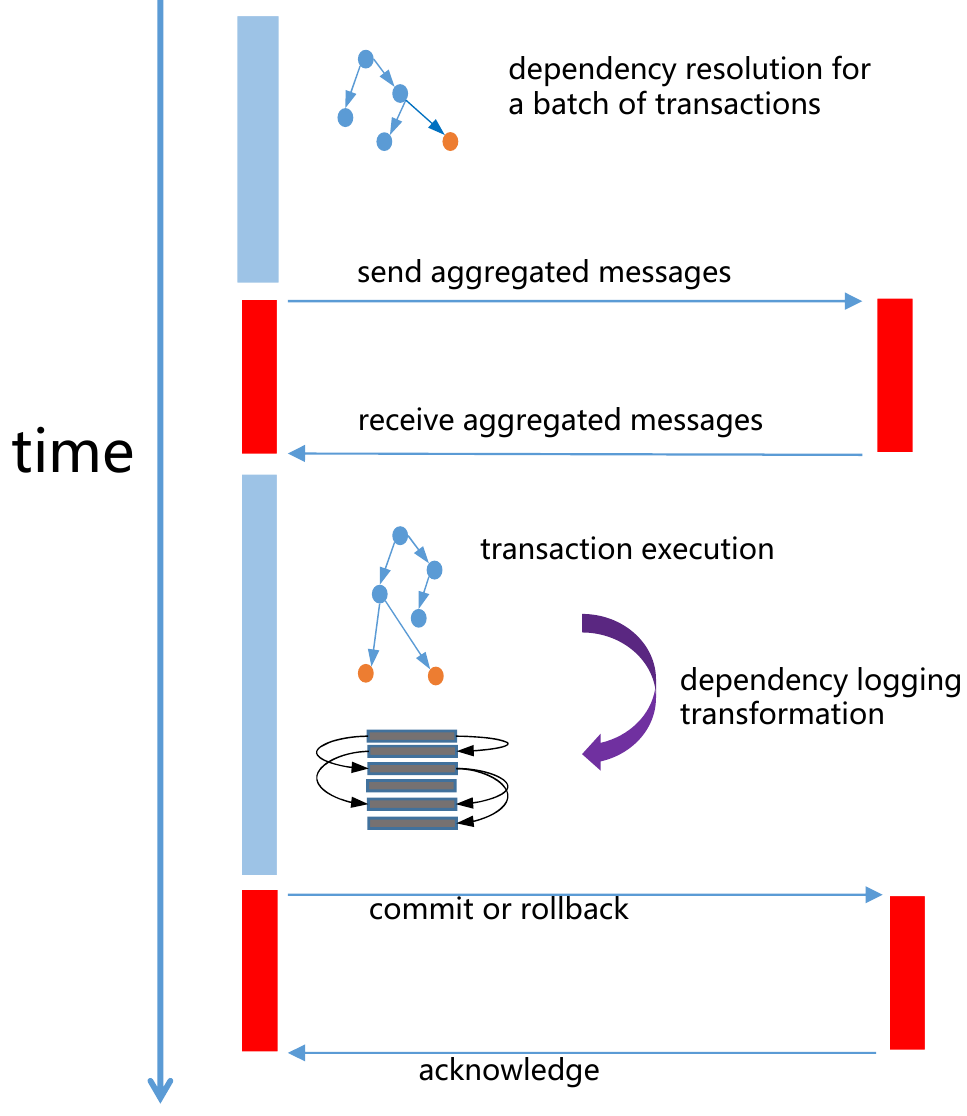}}
	\caption{Life cycle of distributed transaction processing}
	\label{fig:dist_example}
\end{figure}

To address the aforementioned issue, we advocate that \textit{batching} should be treated 
as a first-class citizen in the design of concurrency control for OLTP systems. 
It is a pragmatic consideration for efficient in-memory OLTP in a distributed environment 
based on the following key observations.
First, batch-oriented strategies are already widely adopted in transaction processing to optimize the runtime performance.
For instance, client usually sends a group of requests to severs to optimize network throughput and reduce latency~\cite{towc10:Louie}.
Moreover, as frequent disk I/Os restrict system's performance especially for in-memory systems,
group commit and batch logging become a standard optimization~\cite{sigmod84:DeWitt, sigmodrec15:Tan}.
Second, compared to standalone systems, 
the performance of a distributed OLTP system is mainly affected by distributed transactions which typically incur high coordination and communication cost and impede exploitation of the computation resources, 
as illustrated in Figure~\ref{fig:dist_example}.
With batch processing enabled,
the available computation resources can be better utilized in anticipation.


It has been the practice that the concurrency control and logging are treated as two independent tasks.
This could have been the consequence that data logging has been effective and has been regarded as the de facto protocol for logging needed for recovery.
Moreover, distributed transaction management is more complex than single-node transaction management, 
since the data are distributed and a transaction may access data from multiple nodes.
Transactions are often partitioned based on the data locality so that the number of subtransactions accessing multiple nodes is kept low.  Such information is useful for recovery in the distributed environment.
Therefore, we re-examine the transaction management and recovery problems with the aim of achieving high throughput and fast recovery.

Our overall idea is to exploit the same data structure, the dependency graph, for both concurrency control and logging. 
We first make use of the dependency graph to resolve transaction contentions both within one node and among different nodes,
and derive a set of serializable transaction schedules.
In particular, transactions are executed in discrete temporal batches, each of which is processed through a \textit{dependency resolution phase} followed by
an \textit{execution phase}.
In the \textit{dependency resolution phase}, each transaction is parsed into a set of transaction pieces
and the operation dependency graphs are constructed in parallel.
With the constructed dependency graphs, a set of serializable transaction schedules can be derived.
In the \textit{execution phase}, each transaction schedule is processed by a single thread and thus the overhead due to locks can be eliminated.
While transactions in the same schedule are processed sequentially,
multiple schedules can be processed in parallel with multi-threading and distributed computing to maximize system resource utilization. 
Moreover, the above two processing phases can be conducted in a pipelined manner with respect to the batched transactions.


We note that the dependency graphs derived during transaction processing capture sufficient information for recovery as well. 
We therefore propose a new type of logging strategy, namely \textit{dependency logging}.
Dependency logging reduces the time of log construction by reusing the dependency graphs and hence improves the logging efficiency.
Compared with traditional \aries~\cite{tods92:Mohan}, 
dependency logging captures the dependency information among committed transactions.
Instead of replaying log records in serial order,
parallel replaying is enabled by reconstructing the dependency relations among log records.
Consequently, a bigger degree of parallelism can be exploited and on-demand recovery can be supported when failures occur.

The contributions of this paper are threefold: 
	
\begin{itemize}
\item
We propose Distributed Dependency Graph based Concurrency Control (\distdgcc) that makes use of dependency graphs
to do transaction management in distributed environment.
With the use of the dependency graphs for concurrency control, 
\distdgcc\ facilitates efficient transaction processing and reduces transaction aborts due to conflicts 
by resolving transaction dependencies ahead of transaction execution.
		
\item 
Based on the dependency graphs, we present a new type of logging, dependency logging, which improves logging efficiency
and supports on-demand recovery when system failure occurs.
The novel dependency logging reduces the log construction time and speeds up the recovery process by reusing the dependency graphs constructed during transaction processing.
		
\item 
We conduct extensive experiments to evaluate both \distdgcc\ and dependency logging in distributed environment. 
The results show that \distdgcc\ is efficient and scalable,
and dependency logging is effective for fast recovery upon system failure and even further improves the processing efficiency during the runtime.
\end{itemize}

The rest of the paper is organized as follows. 
Section~\ref{sec:system_design} describes the mechanism of the original Dependency Graph based Concurrency Control (\dgcc) protocol \cite{tkde16:Yao}, 
as well as the corresponding new design for a distributed in-memory OLTP system.
The dependency logging is elaborated on in Section~\ref{sec:fault_tolerance}.
How to do the recovery with dependency logging is discussed in Section~\ref{sec:recovery}.
We conduct an experimental study in Section~\ref{sec:eval} to evaluate both the effectiveness and efficiency of our proposing techniques. 
We review the related work in Section~\ref{sec:related_work} and conclude the paper in Section~\ref{sec:conclusion}.

\begin{figure*}[t]
  \centering
  \includegraphics[scale=0.45]{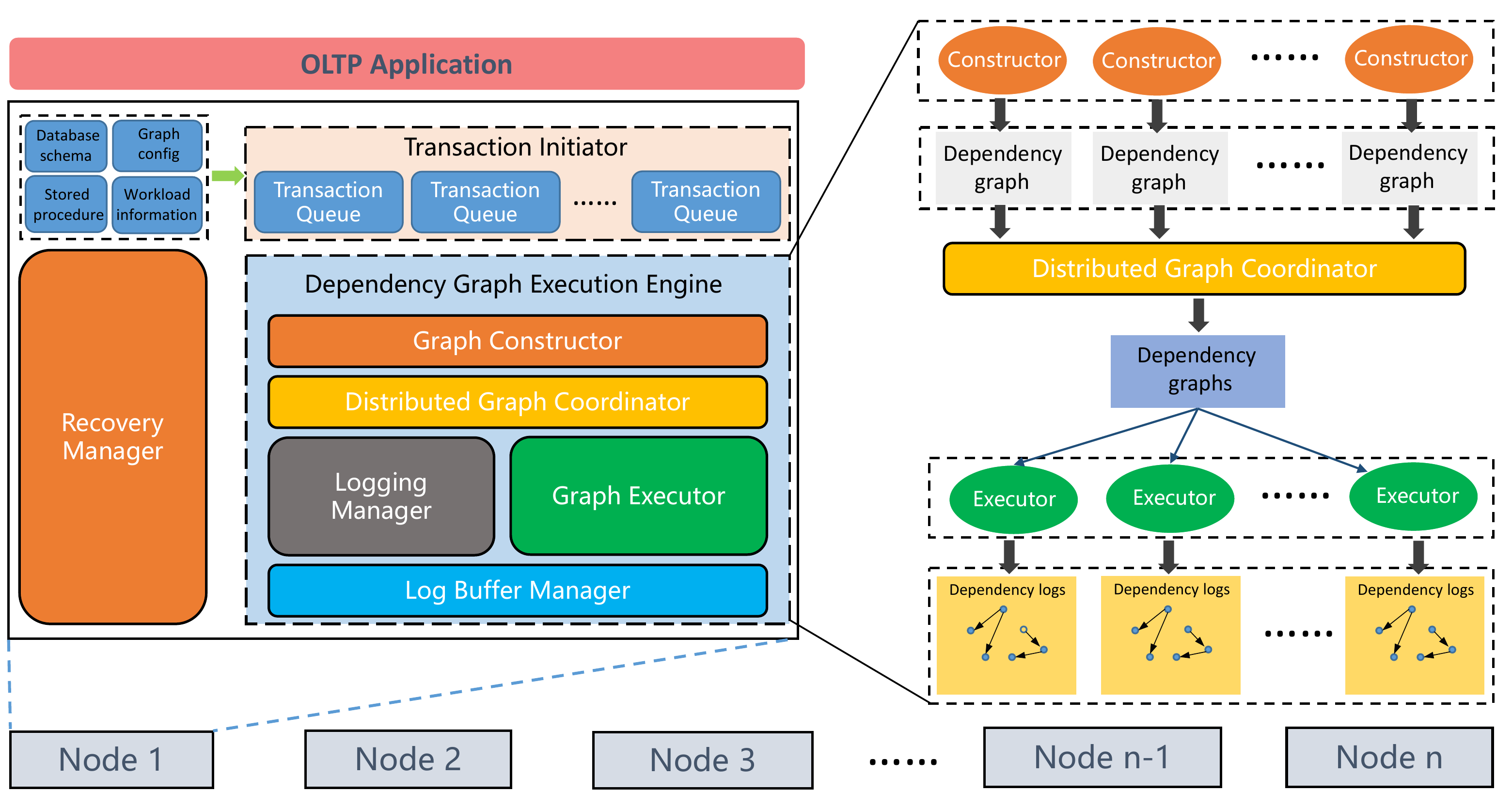}
  \caption{System architecture}
  \label{fig:arch}
\end{figure*}

\section{Background}
\label{sec:system_design}

With the DRAM price decreasing over the recent decade\footnote{\raggedright \url{http://www.anandtech.com/show/10512/price-check-q3-2016-dram-prices-down-over-20-since-early-2015}},
DRAM is replacing disk as the primary storage. 
More and more systems attempt to maintain the whole data in memory to support faster data accesses.
Consequently,
buffer management is no longer a main performance bottleneck for in-memory systems~\cite{vldb08:Kallman}, 
and the efficiency of concurrency control and logging becomes a critical performance issue.
Traditionally, concurrency control and logging have been treated as two separate problems and designed independently, 
which may restrict the computation resource utilization as a consequence.

\noindent\textbf{Concurrency Control:} 
Fundamentally, concurrency control protocols can be broadly classified into two categories: two-phase locking and timestamps ordering.
	 
As a pessimistic protocol,
two-phase locking (2PL)~\cite{cacm76:Eswaran} assumes transactions tend to conflict and 
thus it requires transactions to acquire locks for a particular data record before they are allowed to execute a read/write operation.
By following the 2PL scheme, potential conflicting data accesses can be prevented.
However, the pessimistic concurrency control scheme of 2PL suffers from high overhead associated with synchronization on concurrency meta-data (i.e., locks)~\cite{sigmod16:Ren}.
Moreover, deadlock detection and resolution are also expensive especially in a distributed setting.
As a consequence, optimistic concurrency control (OCC) protocols~\cite{tods81:Kung} based on timestamp ordering are preferred by recent high-performance OLTP systems~\cite{osdi14:Mu}.
The design of timestamp ordering concurrency control schemes is based on the monotonically increasing timestamps, 
which are exploited to process conflicting read/write operations in a proper order. 
%
While OCC exhibits great success with low-contended workloads, 
it performs poorly under high contention due to excessive transaction aborts~\cite{vldb15:Faleiro}.
OCC usually aborts conflicting transactions before the commit time 
and results in wasting precious CPU cycles spent on these transactions that are destined to abort~\cite{tods87:Agrawal}.
Moreover, a centralized timestamps assignment component is usually required in the distributed environment,
which also restricts the system performance and scalability.
By taking a batch of transactions rather a single transaction as the processing unit, 
\dgcc\  shows good performance on both low contention and high contention workloads
by reducing the transaction aborts due to conflicts on multi-core system~\cite{tkde16:Yao}.
It is therefore a natural consideration to extend it to distributed settings.
In Table~\ref{tab:cc_comparison}, we compare 2PL and OCC with \distdgcc\ 
on average latency of distributed transaction and conflicts handling.
As illustrated, both 2PL and OCC take $2$ round-trip time (RTT) 
to commit (or abort) a distributed transaction.
By aggregating messages for a batch of distributed transactions,
\distdgcc\ reduces the network latency significantly.

\noindent\textbf{Logging:} 
A databases system should guarantee that all the changes made by committed transactions are durable and changes from uncommitted transactions are invisible after system recovering from failures~\cite{1989iwdm:agrawal}.
To provide durability, a database system writes the changes of a transaction to log files on durable storage before committing to the client.
We compare \aries, \command\ and Dependency logging in Table~\ref{tab:logging_comparison}.
\aries~\cite{tods92:Mohan} is the most widely used approach in which each log saves one update information on one data record.
Both data images before and after the update are saved in the log.
As a consequence,
\aries\ usually generates large size logs and incurs substantial amount of disk I/Os, which affect the runtime performance especially for in-memory database systems.
\command~\cite{icde14:Malviya} is a coarse-grained logging approach that tracks information about transactions instead of data records and hence reduces the log size.
However, it incurs expensive recovery cost in distributed environment, 
since the whole cluster needs to roll back to the latest checkpoint and replays all the committed transactions in serial order even for a single-node failure.
While \aries\ supports independent recovery for failed nodes,
it still needs to replay log record in committed order.
Consequently, it is hard to achieve high parallelism in recovery compared to its runtime.

\begin{table*}[t]
	\centering
	\begin{tabular}{c|c|c|c} 
	\textbf{Concurrency control} 	& \multirow{2}{*}{\textbf{How to resolve conflicts}}	& \textbf{Distributed transaction's} & \multirow{2}{*}{\textbf{Abort rate}}\\ 
	\textbf{protocols}											&											& \textbf{average commit latency in RTTs} & \\  \hline\hline
		
		2PL      				& Lock	(waiting)					& $2$ 							&	Low\\ 
		OCC						& Abort and retry  					& $2$								&	High for high contention workloads\\ 
		DistDGCC				& Dependency graph					& $2/N$							&	No aborts due to conflicts					\\ 
	\end{tabular}
	\caption{Comparison between \distdgcc, 2PL and OCC. $N$ in the table denotes the number of distributed transactions in one batch for \distdgcc.}
	\label{tab:cc_comparison}
\end{table*}

\begin{table*}[t]
	\centering
	\begin{tabular}{c|c|c|c} 
	\textbf{Logging approach}	 	& \textbf{Logging content}			&	\textbf{Runtime cost}	& \textbf{Recovery}							\\ \hline\hline
		\aries					& Data record images		& High		&	Failed node rollback and serial replaying		\\
		\command				& Transaction information & Low 		&	Cluster rollback and serial replaying \\
		Dependency Logging		& Operation and dependency information		& Low & Failed node rollback and parallel replaying\\
	\end{tabular}
	\caption{Comparison among Dependency logging, \aries\ and \command.}
	\label{tab:logging_comparison}
	\vspace{-1mm}
\end{table*}

\subsection{DGCC Overview}
\label{sec:dgcc}

By taking the batching technique as the first-class citizen in concurrency control design, 
dependency graph based concurrency control (\dgcc)~\cite{tkde16:Yao} which facilitates lock-free transaction processing and reduces transaction aborts caused by contention is proposed for multi-core systems.
By separating the contention resolution from the transaction execution, \dgcc\ processes transactions in discrete temporary batches.
Each batch is processed through a contention resolution phase followed by an execution phase.
 	
By parsing the statements of a transaction, 
worker thread resolves the dependency relations within one single transaction and decomposes it into a set of \textit{record actions}.
Record action is an abstraction of consecutive operations that are conducted on the same data record within one transaction.
In the contention resolution phase, 
each worker thread maintains a constructor to resolve the dependency relations as \textit{Logical Dependency} or \textit{Temporal Dependency} 
among a batch of transactions and builds the dependency graph accordingly.
Logical dependency determines the logical execution order of record actions within one transactions 
and temporal dependency determines the execution order of conflicting record actions from different transactions.	
This process is fast and efficient as all the work are done by the single-threaded model~\cite{vldb08:Kallman}.
To better exploit the hardware parallelism, 
multiple dependency graphs are constructed in parallel by different worker threads that are bound to distinct CPUs. 
A synchronization operation is performed to guarantee that all worker threads finish the dependency graph construction before the execution phase is initiated.
The dependency graph partitions are evenly distributed to available worker threads and the execution can therefore be conducted in parallel.

\begin{figure*}[ht]
  \centering
  \subfigure[Dependency relations in local node]{
    \label{fig:running_example_1}
    \includegraphics[height=0.3\linewidth, width=0.45\linewidth]{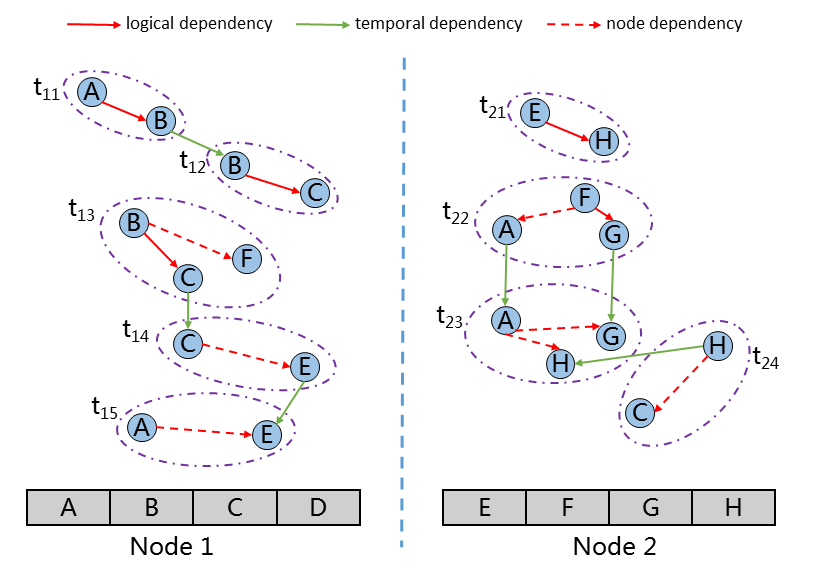}}
  \subfigure[Global dependency relations]{
    \label{fig:running_example_2}
    \includegraphics[height=0.3\linewidth, width=0.45\linewidth]{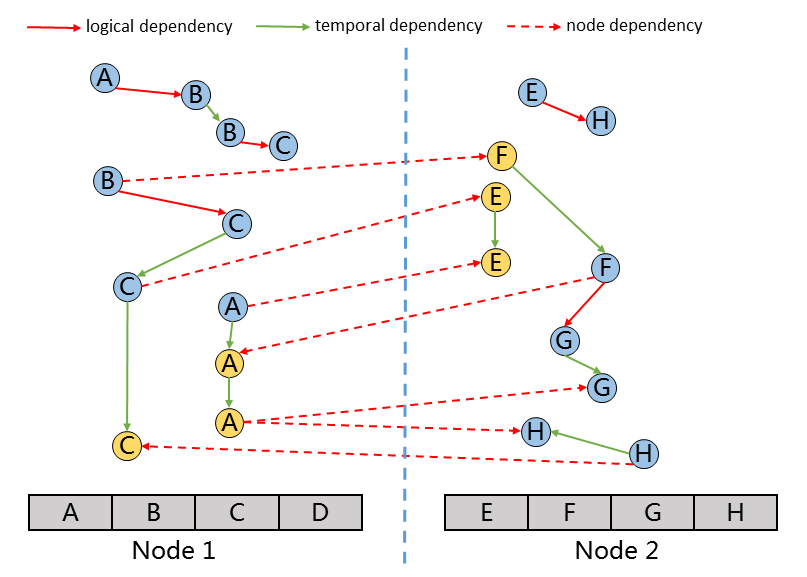}}
  \caption{A running example}
  \label{fig:running_example}
\end{figure*}

\subsection{Distributed \dgcc}
\label{sub:distdgcc}

While the aforementioned \dgcc\ protocol fits the multi-core and main-memory architecture well, 
it is nontrivial to apply to the distributed setting.
The main challenges are twofold:
(1) The original \dgcc\ constructs dependency graphs in parallel for several batches of transactions. 
However, dependency graphs should be executed in a serial order.
When the cluster is large, the graphs executed latter may have to wait for a long time before being executed, 
which may increase the latency dramatically.
(2) While node failures are infrequent in reality, an increasing number of nodes in distributed environment probabilistically leads to more failures and results in expensive recovery cost~\cite{fast2007:pinheiro,nordac2004:roos,tdsc2010:schroeder}. 
Thus, an efficient recovery mechanism that supports \dgcc\ is essential.
Therefore, instead of treating both concurrency control and recovery as two distinct problems, we make use of the information used in \dgcc\ in logging for supporting fast recovery.
However, for ease of explanation, we shall first describe our concurrency control protocol for distributed transaction management, followed by recovery based on dependency logging.

We therefore first propose \distdgcc, a transaction management solution specifically designed for distributed environment.   
\distdgcc\ introduces a coordination mechanism to adjust the execution order of distributed transactions,
which not only guarantees the serializability, but also facilitates high execution parallelism.
By aggregating network messages for one batch of transactions,
\distdgcc\ reduces the thread waiting time for network blocking and thus optimizes the average latency.
Since local transactions are executed differently from distributed transactions, we distinguish the two below.

\noindent \textbf{Local Transaction:} 
It accesses data accessed stored in a single compute node.
For a local transaction $T_{local}$ running on one node, 
we add $T_{local}$'s corresponding vertices and edges into the dependency graph constructed in the node 
by directly following the protocol as described in~\ref{sec:dgcc}.
Note that both $T_{local}$'s intra-transaction dependencies and inter-transaction dependencies related to $T_{local}$ 
are automatically resolved by the resulting dependency graph. 

\noindent \textbf{Distributed Transaction:} 
it accesses data stored in multiple compute nodes.
For a distributed transaction $T_{dist}$ whose execution spans multiple compute nodes, 
we add $T_{dist}$'s corresponding vertices and edges into the dependency graphs of the related nodes. 
Although the inter-transaction dependencies related to $T_{dist}$ are resolved by the resulting dependency graphs, 
$T_{dist}$'s intra-transaction dependencies among its inter-node actions are not captured by the dependency graphs due to the distribution. 	
To address this issue, 
we define an additional type of dependency, namely \textit{Node Dependency}, to track the dependencies among inter-node actions of the same transaction.

\vspace{-2mm}
\begin{definition}{Node Dependency.}
A node dependency exists between vertex $v_i$ and vertex $v_j$, denoted as
	$$v_i \succ_{\mbox{\scriptsize dist}} v_j$$
if and only if $v_i \succ_{\mbox{\scriptsize logical}} v_j$ and $v_i$, $v_j$ are executed in different compute nodes.	
\end{definition}

In a distributed system, it is important to take advantage of data locality to optimize system's overall performance.
The ideal case is that each compute node works independently.
Fortunately, batching is considered as a high priority in \distdgcc.
It is easy to separate local transactions from distributed transactions during the dependency graph construction.
In our implementation, local transactions in each batch are executed independently following the original \dgcc\ 
protocol~\cite{tkde16:Yao} to achieve high parallelism both among cluster and in a single node.
For distributed transactions with node dependency, 
\distdgcc\ distributes the vertices to relevant nodes according to the data locality in a batch manner, 
which reduces the number of network messages. 
As shown in Figure~\ref{fig:arch}, we introduce a \textit{Distributed Graph Coordinator} in each node to help distribute and receive vertices with dependency information to relevant compute nodes.
Along with this information, some meta-data (e.g., node ID) are also sent to resolve temporal dependency among vertices from different compute nodes.
After receiving all vertices in one batch, the Distributed Graph Coordinator is responsible for constructing a new dependency graph for those vertices.

Figure~\ref{fig:running_example_1} illustrates a running example on two compute nodes.
A batch of transactions $t_{11}$, $t_{12}$, $t_{13}$, $t_{14}$, $t_{15}$ runs on the node $1$.
$t_{11}$ and $t_{12}$ are local transactions and the rest are distributed transactions.
During the contention resolution phase, each node constructs the dependency graphs for local transactions and distributed transaction respectively.
The dependency graph of all local transactions contains logical dependency and temporal dependency.
Nodes are not required to communicate with each other for the execution of local transactions.
However, node dependency may exist in distributed transactions that cannot be executed locally.
Taking $t_{13}$ as an example, it reads record $B$ and updates records $C$ and $F$. 
The challenge is that $t_{13}$ does not know any information about $F$, and thus it is hard to resolve the conflicts among transactions from other nodes.    
Therefore, before the transaction execution, the Distributed Graph Coordinator partitions the original dependency graph along the node dependency edges and distributes subgraphs to relevant nodes.
As shown in \ref{fig:running_example_2}, node $1$ sends the yellow vertices along with their relevant edges to node $2$.
The Distributed Graph Coordinator on node $2$ collects the subgraph and detects the temporal dependency relations between the local subgraphs.
Then it builds a global dependency graph for all distribute transactions that touch data on node $2$.

\begin{figure*}[t]
  \centering
  \includegraphics[width=0.98\linewidth, height=0.06\linewidth]{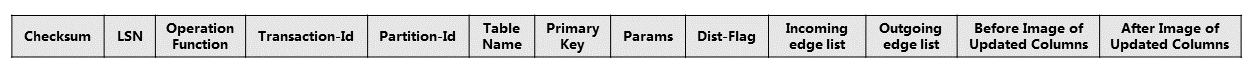}
  \vspace{-2mm}
  \caption{Fine-grained dependency logging record structure}
  \label{fig:dependency_logging_record_structure}
\end{figure*}

\section{Dependency Logging}
\label{sec:fault_tolerance}

Fault tolerance and recovery are important to database system which should guarantee the ACID (Atomicity, Consistency, Isolation, Durability) properties.
We adopt the log-based fault-tolerant scheme, which is a tried-and-tested approach in database literature.
Operationally, it generates transaction logs at runtime and performs recovery upon system failure based on the logs. 
Specifically, each transaction saves the recovery-oriented information into logs along its execution and then flushes the logs to persistent storage before it commits.
When system failure occurs and the recovery process is launched, 
partially processed transactions need to be undone and committed transactions need to be redone according to the materialized logs.
Generally, two aspects of performance are concerned in the design of log-based fault tolerance:
one is for runtime logging, and the other is for failure recovery.
On the one hand, transactions that construct their logs at runtime definitely introduce performance overhead,
which affects the system throughput and processing latency.
Therefore, optimized designs aim to make runtime logging as efficient as possible. 
On the other hand, undoing and redoing transactions during failure recovery typically incur significant system downtime. 
Towards better system utilization, 
the impact of system downtime ought to be restricted in terms of the affected nodes and the checkpointing interval.
	
While the high-level idea of log-based fault-tolerant scheme for distributed OLTP systems is straightforward, 
efficiently supporting low-overhead logging at runtime and fast recovery upon system failure is non-trivial, 
especially with the traditional fault-tolerant schemes, e.g., \aries~\cite{tods92:Mohan}.
\aries\ is widely adopted and is considered a ``heavy-weight" logging approach, 
as it generates fine-grained logs to describe how data records have been changed by the transactions.
Since all the recovery-oriented information is stored in the log records, 
its log structure is complicated, which usually takes more CPU cycles to generate.
Also, more disk I/Os are required to flush the logs to disk, further affecting the system runtime performance.
While \aries\ supports independent recovery when system failure happens, 
it is hard to exploit parallelism during the recovery due to its sequential log processing.
	
To maximize the performance gain of \distdgcc\ and optimize failure recovery process, 
a new type of \textit{dependency logging} is proposed based one dependency information generated during transaction processing.
Dependency logging is a variant of write-ahead logging. 
It not only logs how the data are changed, but also logs the dependency information among the updated data.
Our system considers transaction management and fault-tolerance within the same framework. 
Specifically, the dependency graphs that are constructed during transaction processing can be reused and transformed to the log records with little overhead.
The dependency graphs used for concurrency control is at record-level, which may generate a large amount of log data.
To further improve the logging efficiency, we then propose a coarse-grained optimization.
We distinguish them as \textit{fine-grained dependency logging} and \textit{coarse-grained dependency logging}, respectively, and detail them in subsections that follow.


\subsection{Fine-grained Dependency Logging}
\label{subsec:dplogging-f}
Traditional logging schemes usually generate log records during the runtime that incurs additional cost to create  and maintain log data structures.
In order to generate log records more efficiently,
fine-grained dependency logging reuses the dependency graphs derived for concurrency control at the runtime.
The data structure of log record written for each record action is shown in Figure~\ref{fig:dependency_logging_record_structure}.
Most of the fields in the log structure are also maintained in the dependency graphs, e.g. the incoming and outgoing edge list.
As a consequence,
the fine-grained dependency logging saves many CPU cycles for generating log records.


\begin{figure}[h]
	\subfigure[\aries]{
		\label{fig:dplogging_aries}
		\includegraphics[scale=0.29]{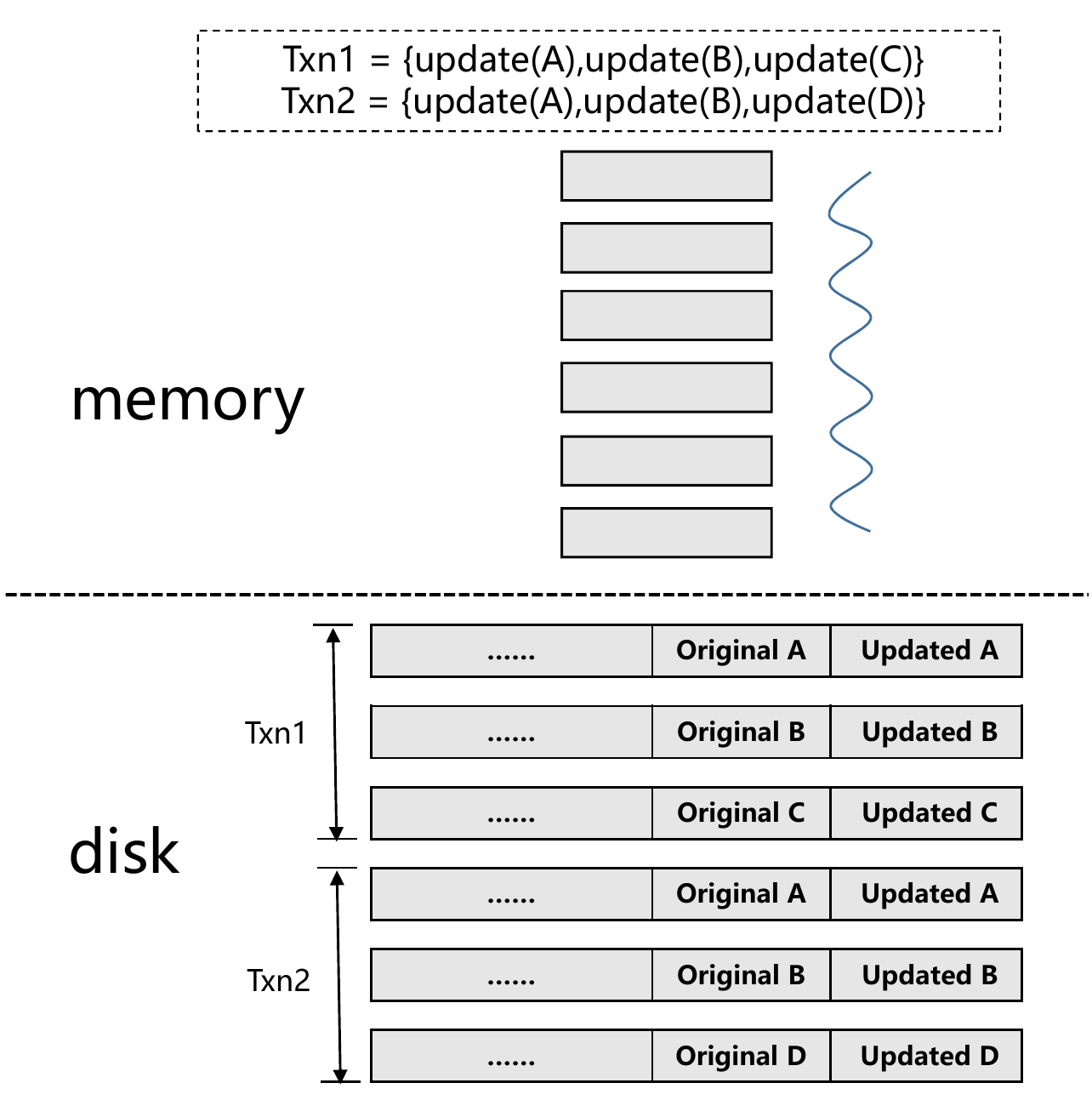}}\quad
	\subfigure[Dependency Logging]{
		\label{fig:dplogging_dp}
		\includegraphics[scale=0.29]{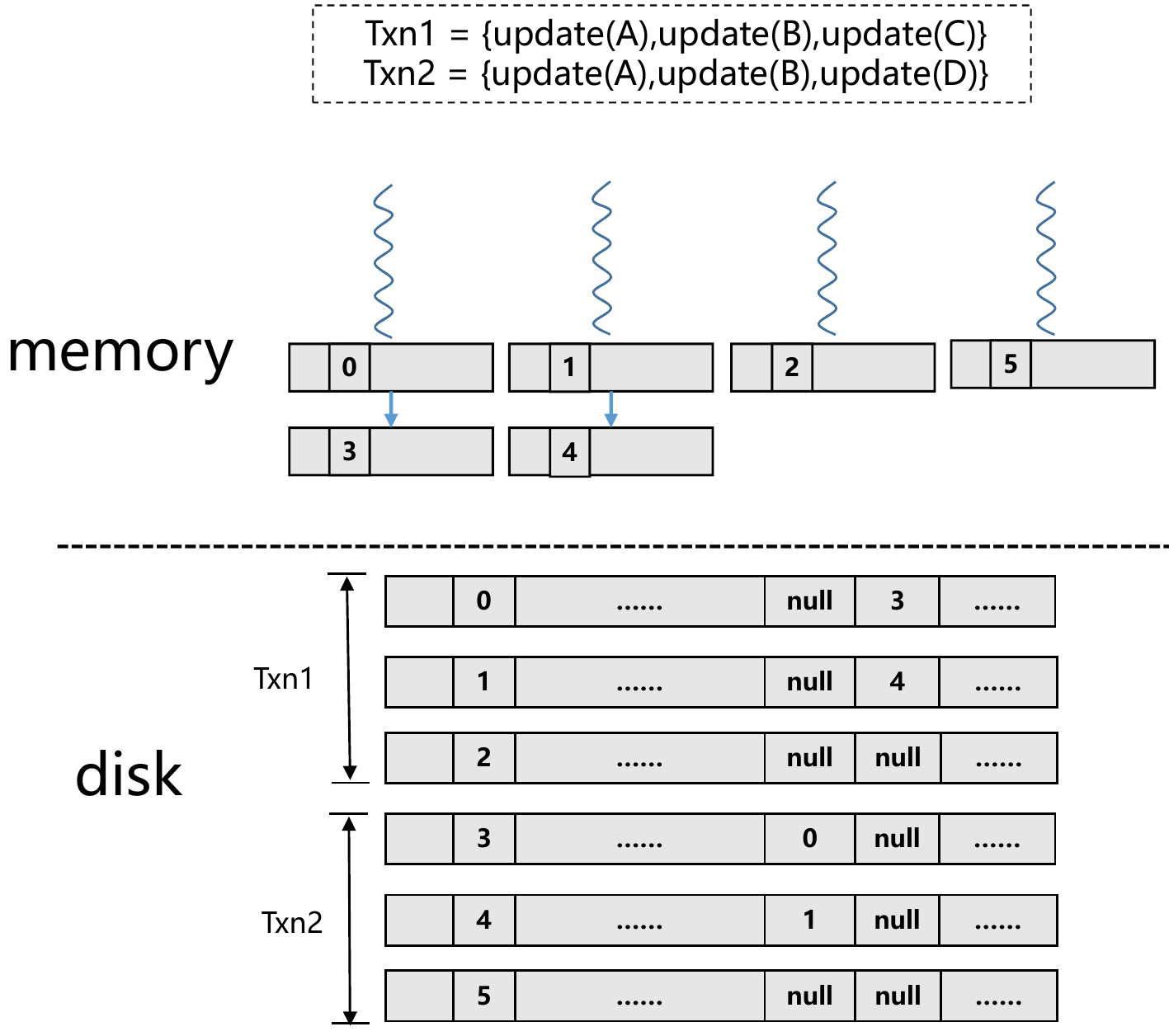}}
	\caption{Example of dependency logging}
	\label{fig:dlogging_example}
\end{figure}

Fine-grained dependency logging writes a unique logical sequence number (LSN) for each log record.
An active transaction table is maintained on each node to help track the LSN of the last flushed log record for each transaction.
Besides, each log record also contains the operation function name and its relevant parameters, 
with which system can restore the data record to a correct state.
Each fine-grained dependency logging record maintains the change information for one data record that can be uniquely referred by the (\textit{table-name, primary-key}) pair.


In each log record, fine-grained dependency logging also keeps track of the dependency information, 
incoming and outgoing edges, with which our system can support on-demand recovery 
and further improve the recovery parallelism on the failed node.
As shown in Figure~\ref{fig:dlogging_example}(a), 
when failure occurs, \aries\ first loads the logs from disk to memory and redoes log records generated by committed transaction in sequence.
It is hard for \aries\ to exploit more parallelism in the redo process, 
since there is little information that can be used to resolve the possible conflicts.
In this example, \aries\ has to redo the six log records in order to recover the database to the correct state.
By saving the dependency information in each log record,
fine-grained dependency logging first loads the log records from disk to memory and rebuilds the dependency structures among log records.
Then the redo process can be executed in parallel.
As the example shown in Figure~\ref{fig:dlogging_example}(b),
four threads can work in parallel to finish the recovery.
Moreover, fine-grained dependency logging can further reduce the downtime compared to other logging approaches.
For example, system with \aries\ cannot start to execute new transactions before the failed node is fully recovered.
By rebuilding the dependency relations for the committed transactions in advance, 
system with fine-grained dependency logging can execute newly arrived transaction immediately.
The newly arrived transaction first recovers all its dependent data and then does the execution, which reduces the system downtime.
	
The field \textit{Dist-Flag} in fine-grained dependency logging's record structure indicates its locality attribute
and it distinguishes the fine-grained dependency logging records into two classes: 
local dependency logging record and remote dependency logging record.
The log records produced by local transaction are all local dependency logging records.
For a distributed transaction that has a coordination node, 
the log records maintained on the coordination node are local dependency logging records 
and the rest are remote dependency logging records.
Like \aries, the remote dependency logging record also stores the data images before and after the change,
with which the dependency relation among different nodes are resolved.
While this design increases the size of remote dependency logging record,
it improves the recovery efficiency. 
More details are discussed in Section~\ref{sec:recovery}.

\subsection{Coarse-grained Dependency Logging}
\label{subsec:dplogging-c}
\begin{figure}[h]
  \includegraphics[scale=0.23]{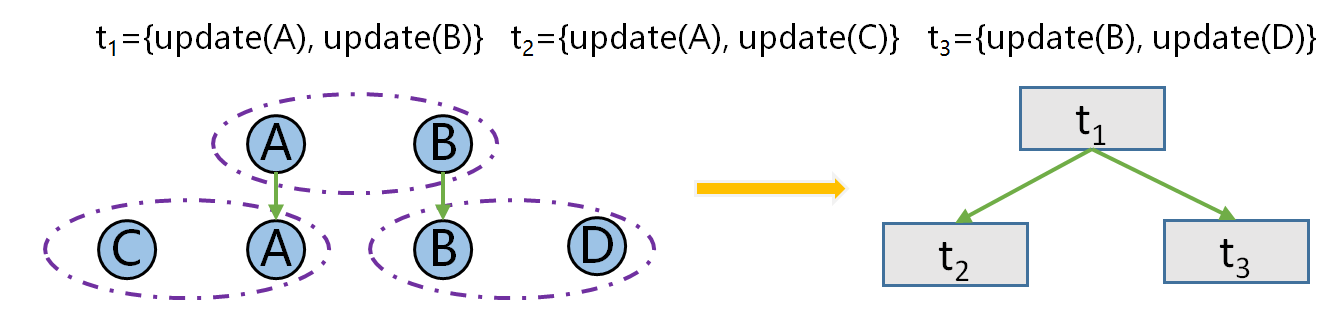}
  \caption{Dependency graph transformation}
  \label{fig:txn_dependency}
\end{figure}

While fine-grained dependency logging increases parallelism and improves the efficiency during the recovery, 
it generates record-level log records that also increase the log size, and thus incurs a substantial number of disk I/Os.
Consequently, coarse-grained dependency logging is proposed to reduce the log size and improve the logging efficiency.
Instead of tracking the dependency information among updated data records, 
coarse-grained dependency logging tracks dependency information among transactions by parsing the record-level dependency graphs.
Intuitively, it needs to traverse the whole dependency graph to complete the transformation, 
which restricts the performance, especially when the dependency graph is large.
Instead, our system does the transformation during the dependency graph construction.
Specifically, if a temporal dependency edge is inserted between record actions $\alpha_i$ and $\alpha_j$ where $\alpha_i \in t_p$ and $\alpha_j \in t_q$, 
an edge should also added between $t_p$ and $t_q$ in the transaction dependency graph.
As shown in Figure~\ref{fig:txn_dependency},
in the dependency graph construction phase, there are three transactions $t_1$, $t_2$ and $t_3$.
When $t_2$ is parsed, a temporal dependency edge should be inserted into the dependency graph.
Similarly, there should be an edge from $t_1$ and $t_2$ in the transaction dependency graph. 

\begin{figure*}[t]
  \includegraphics[width=0.98\linewidth, height=0.06\linewidth]{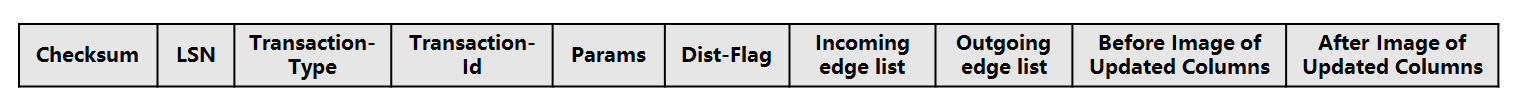}
   \vspace{-2mm}
  \caption{Coarse-grained dependency logging record structure}
	\vspace{-1mm}
  \label{fig:coarse_grained_structure}
\end{figure*}

The coarse-grained dependency logging simply tracks how transaction works and all its dependency relations.
Besides the transaction information, each coarse-grained dependency logging record also contains the dependency information among other transactions.
Each log record that is written out for each transaction has the structure as shown in \ref{fig:coarse_grained_structure}.
Like the fine-grained dependency logging record, there is also a \textit{Dist-Flag} field here that indicates whether the transaction is distributed transaction or not.
For all local transactions, the fields ``Before Image of Updated Columns'' and ``After Image of Updated Columns'' are empty.
Otherwise, the changed data image should be saved in the log record.
Unlike traditional \command\ that only stores a single log entry for distributed transaction (usually on its coordination node), 
coarse-grained dependency logging creates log entries on all its participating nodes.
Instead of recording all the changes of the distributed transaction, each log record only saves recovery-oriented information where it locates. 
By doing this, dependencies among nodes can be resolved and independent recovery can be achieved.
Compared to fine-grained dependency logging, 
coarse-grained dependency logging not only keeps all its strengths but also reduces the log size, e.g., the number of log record in Figure~\ref{fig:dlogging_example} can be reduced to $2$.

\subsection{Dependency Log Recording}

As discussed in Section~\ref{sub:distdgcc},
our system executes local transactions and distributed transactions separately even in one batch of transactions.
It is simple to write the dependency log records for local transactions, since each node does the execution independently.
When all the local transactions are processed, system writes all the dependency log records to the disk at one time to maximize the performance.
As all data can be fetched locally, 
each dependency logging record mainly saves dependency information, 
specifically the incoming and outgoing edges either among log records or transactions. 
For distributed transaction,
the stored information differs for local dependency logging records from remote dependency logging records.	
Both local and remote log records write incoming and outgoing edges.
For remote dependency logging record, it also writes the data images before and after the change.
	
To further improve the performance of our proposed dependency logging,
some optimizations are adopted.

\subsubsection{Log Compression}
\label{subsec:log_compression}
Since the dependency graphs constructed during the runtime are directed graphs, 
it is sufficient to store such edge information in one column instead of two columns.
Like the example shown in Figure~\ref{fig:dlogging_example}(b), 
the dependency logging record $0$ contains the information that its outgoing edge is to record $3$.
Record $3$ also knows that its incoming edge is from record $0$.
Hence, dependency logging can remove the field ``Incoming edge list'' from the log structure to reduce the log size.
	
To further improve the recovery efficiency, 
the remote dependency logging record contains the data images before and after the change.
For tables with wide rows, it wastes a huge amount of log space to save the data images of the whole record, 
especially when the transaction only changes a small set of columns.
Thus, the log size can be further reduced by indicating which columns are changed by the transaction.
Then only those changed columns are saved in the log record instead of the whole row.

\subsubsection{Batch-Oriented Optimizations}
\label{subsec:batch_optimization}

Group-commit that groups multiple log records and flushes to disk at one time is already widely adopted.
It reduces the number of disk writes and thus improves the logging performance.
However, systems that execute transactions individually usually write log data to a shared log file,
simplifying the recovery at the expense of higher logging overhead.
In \distdgcc, transactions are executed in a batch manner
and both fine-grained and coarse-grained dependency logging rely on dependency information rather than commit sequence to perform the recovery.
Hence, each worker thread maintains an individual log buffer to avoid lock contention.
%

\section{Recovery}
\label{sec:recovery}
	
When a system performs failure recovery, 
system consistency and availability are the two main concerns.
While traditional logging approaches (e.g., \aries) have proven guarantee of durability, 
systems using such logging approaches inevitably incurs longer system downtime during recovery, 
which significantly degrades the system availability (even replication-based approaches provide higher availability in database cluster, they still suffer for whole cluster failure).
In contrast, the proposed dependency logging not only guarantees the durability, 
but also provides a better availability during recovery.
We justify the above claim by first looking into how single-node failure is handled under the dependency logging scheme, 
and then extending the discussion to the case of cascading failure. 

\subsection{Recovery of Single-Node Failure}
\label{subsec:single_node_recovery}
	
\begin{figure}[h]
	\centering
	\subfigure[Fine-grained dependency logging records]{
		\centering
		\label{fig:recovery_dpf}
		\includegraphics[scale=0.5]{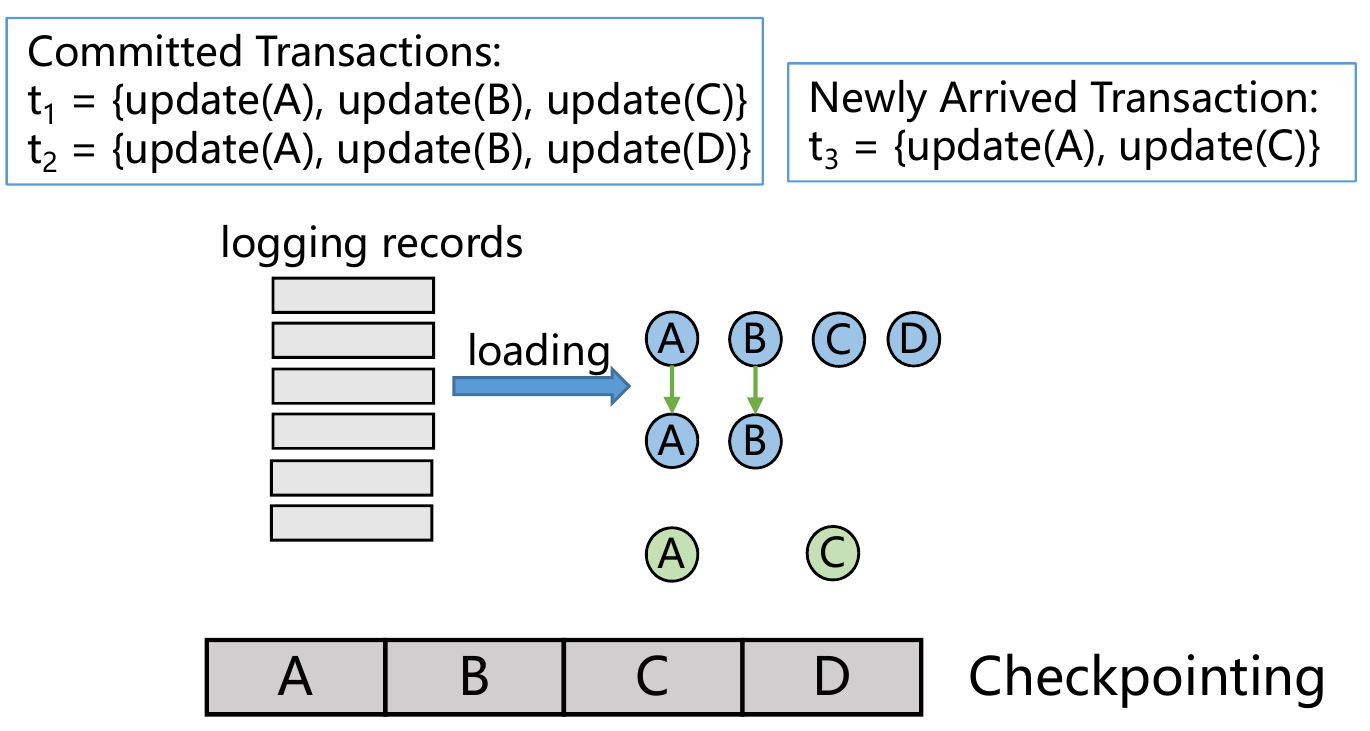}}\quad
	\subfigure[Coarse-grained dependency logging records]{
		\centering
		\label{fig:recovery_dpc}
		\includegraphics[scale=0.5]{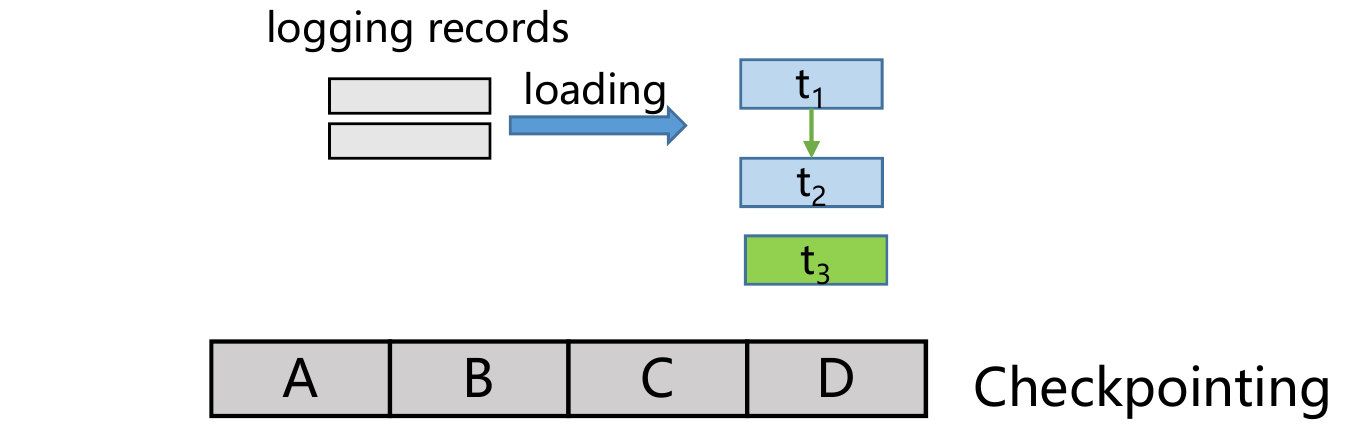}}
	\caption{Single-node recovery from dependency logging}
	\label{fig:recovery_example_1}
\end{figure}

The recovery for the dependency logging starts by reloading the latest database snapshot from the disk. 
After which, the system reloads log files and restores necessary data structures (e.g., indexes).
Since the dependencies among committed transactions are tracked by the dependency logging, 
redoing these transactions can be performed by replaying their execution based on the corresponding rebuilt dependency graphs.
Compared to coarse-grained dependency logging in which each committed transaction has only one log record, 
fine-grained dependency logging records of a transaction may not be fully written to the disk.
Thus, system needs to first remove the log records from the incomplete transaction before starting the recovery process, 
which is like the undo operation for \aries. 
As such recovery process is equivalent to the normal execution phase, 
system can still accept new transactions during recovery. 
That is, the redoing transactions form a batch which is processed first, 
and the newly arrived transactions form more batches which will be processed subsequently. 
However, this does not lead to any improvement on system availability,
since newly arrived transactions cannot be executed before all committed transactions are totally redone.
Instead, the system executes newly arrived transactions sequentially (i.e., enforcing batch size to be $1$).
In order to guarantee a consistency, the new arrived transaction executes only after all its dependent data are recovered.
As shown in Figure~\ref{fig:recovery_example_1}, 
node $1$ encounters a failure after $t_1$ and $t_2$ commit. 
By recovering data record $A$ and $B$ to the correct state, 
the newly arrived transaction $t_3$ can be executed without waiting for the whole database to be fully recovered.
This on-demand recovery mechanism enables normal transaction processing and recovery work in parallel, 
which reduces the downtime for failure and improves the system availability.
When the failed node is fully recovered,
The system increases the batch size that optimizes the overall performance.


The example in Figure~\ref{fig:recovery_example_1} illustrates the trade-off between fine-grained and coarse-grained dependency logging.
While coarse-grained dependency logging usually achieves better runtime performance due to smaller log size, 
it sacrifices some degree of parallelism for the recovery and usually leads to a higher latency for transaction that is processed during the recovery.

\begin{figure*}[t]
	\centering
	\subfigure[YCSB workload with low contention]{
		\label{fig:ycsb_local_5} 
		\includegraphics[scale=0.3]{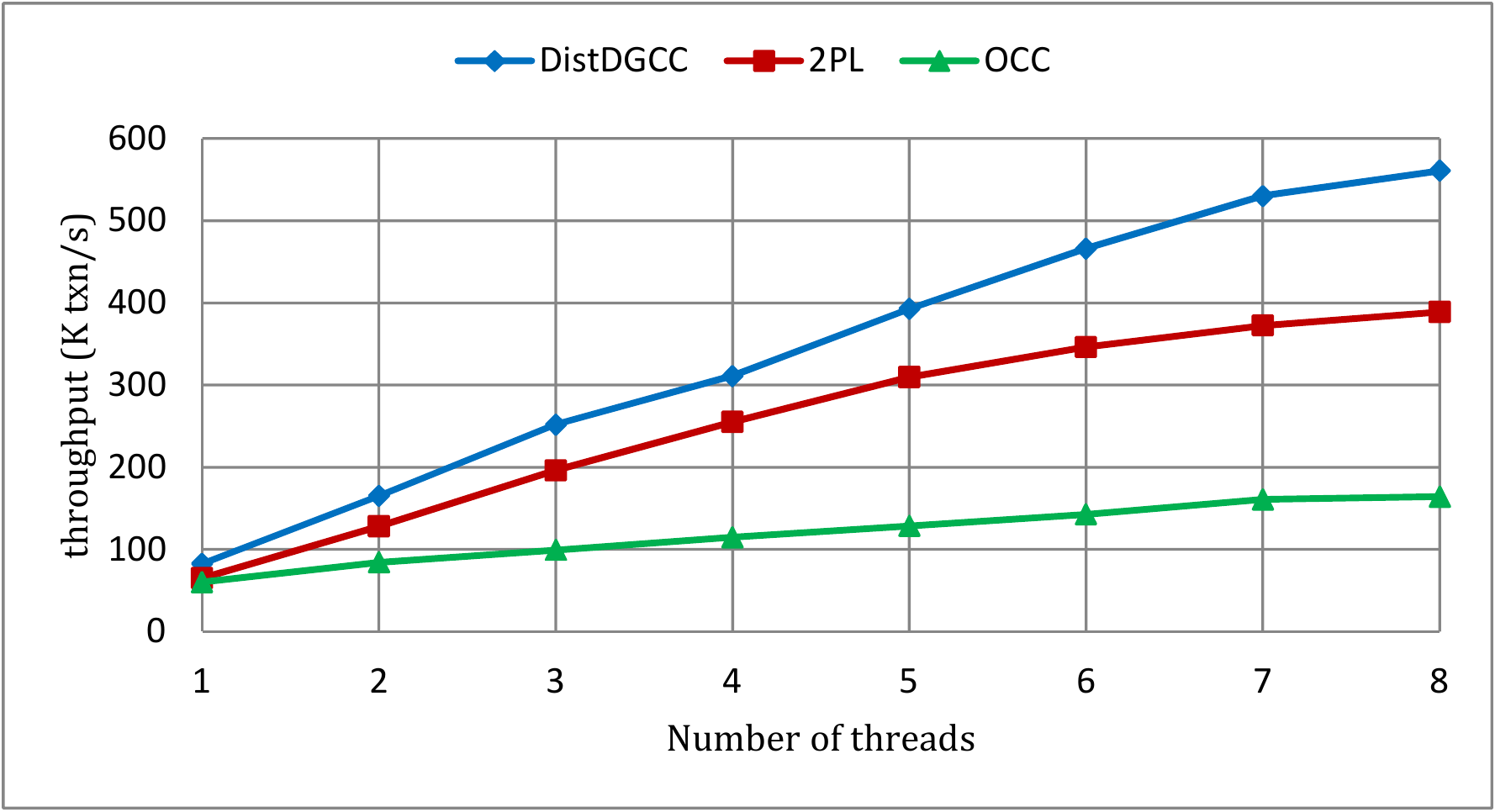}}\quad
	\subfigure[YCSB workload with high contention]{
		\label{fig:ycsb_local_8} 
		\includegraphics[scale=0.3]{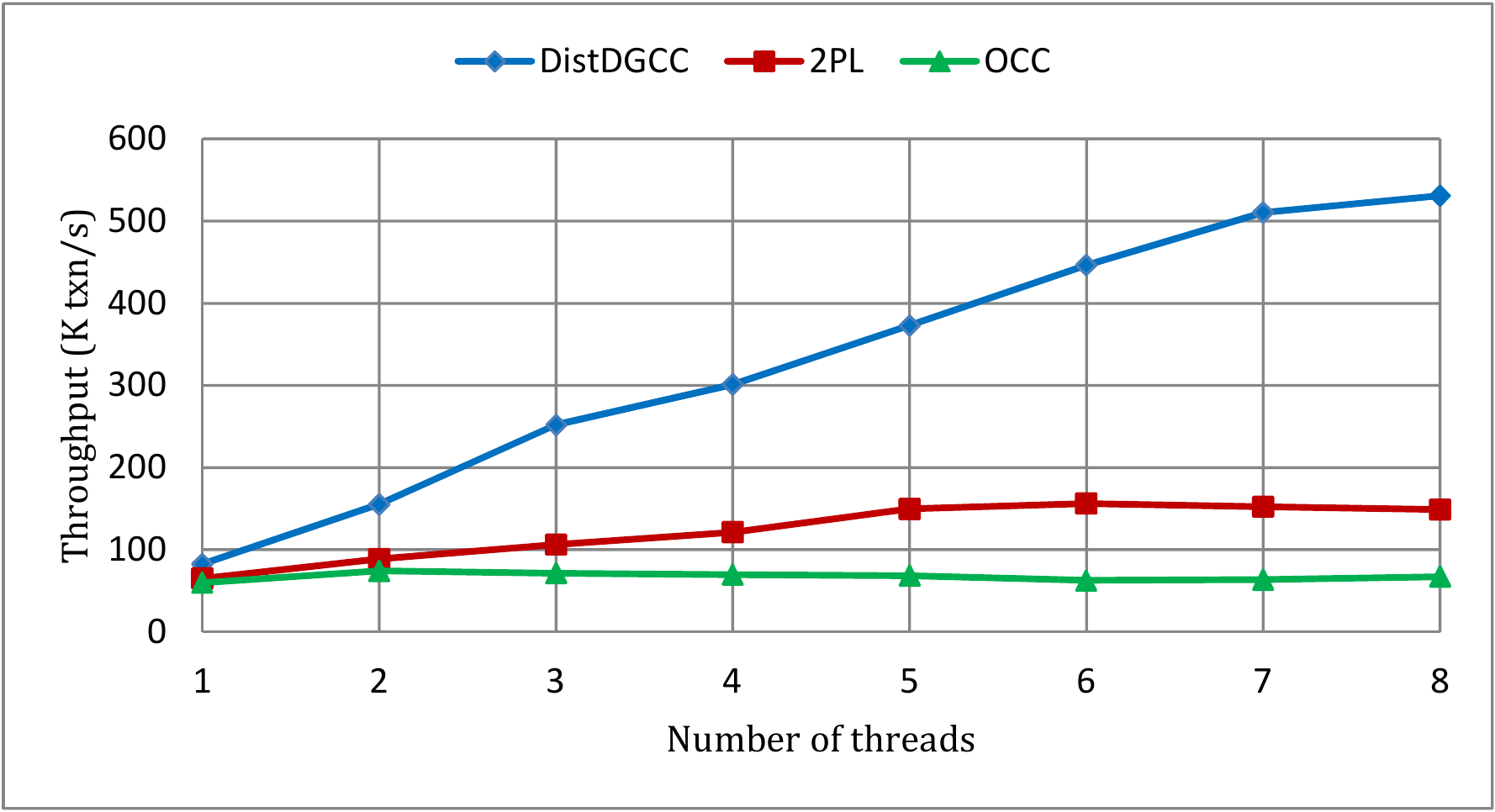}}\quad
	\subfigure[TPC-C workload]{
		\label{fig:tpcc_local} 
		\includegraphics[scale=0.3]{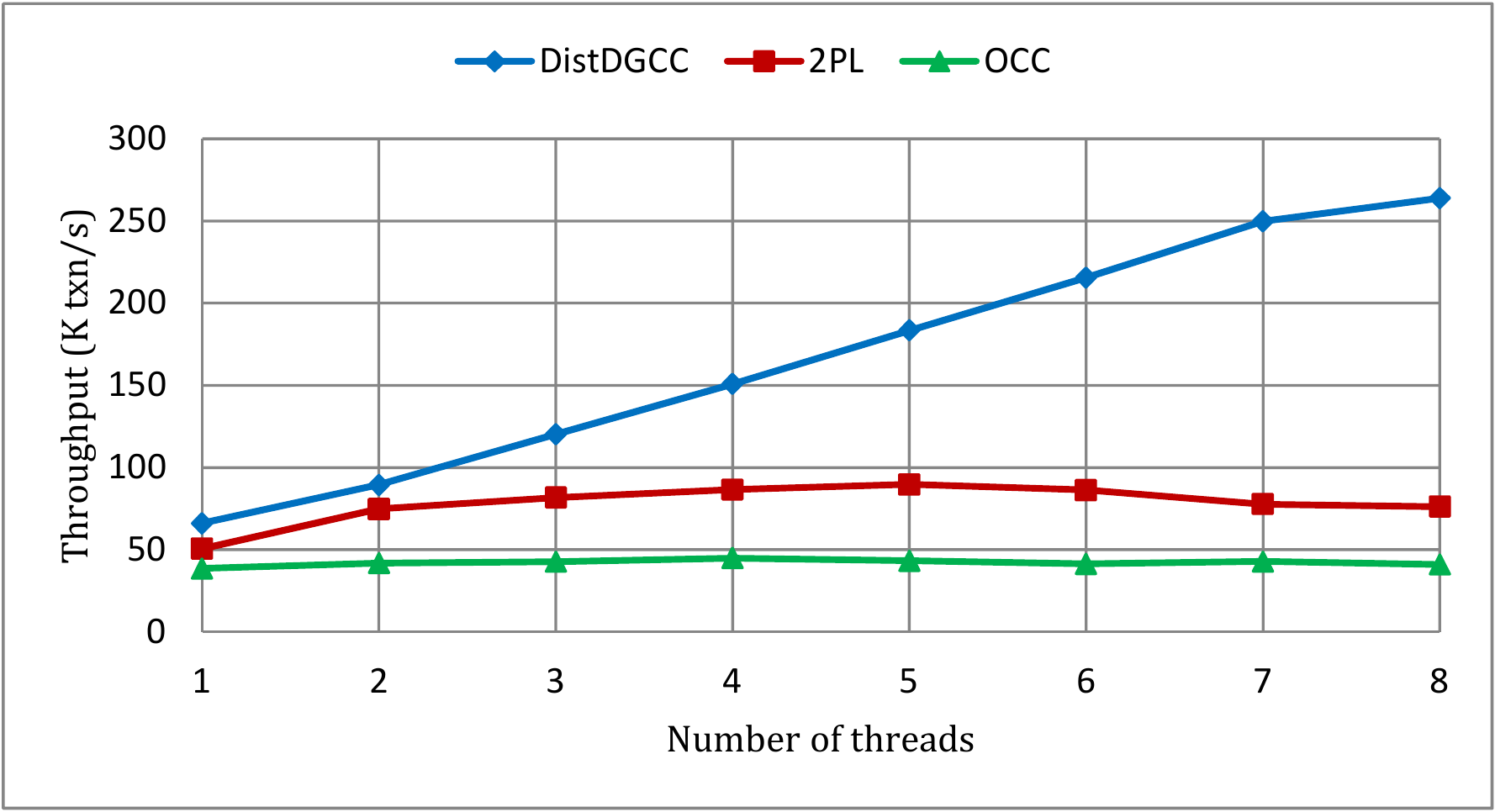}}
	\caption{Throughput evaluation}
	\label{fig:distdgcc_local_thr}
\end{figure*}

\begin{figure*}[t]
	\centering
	\subfigure[YCSB workload with low contention]{
		\label{fig:ycsb_local_5_lat} 
		\includegraphics[scale=0.3]{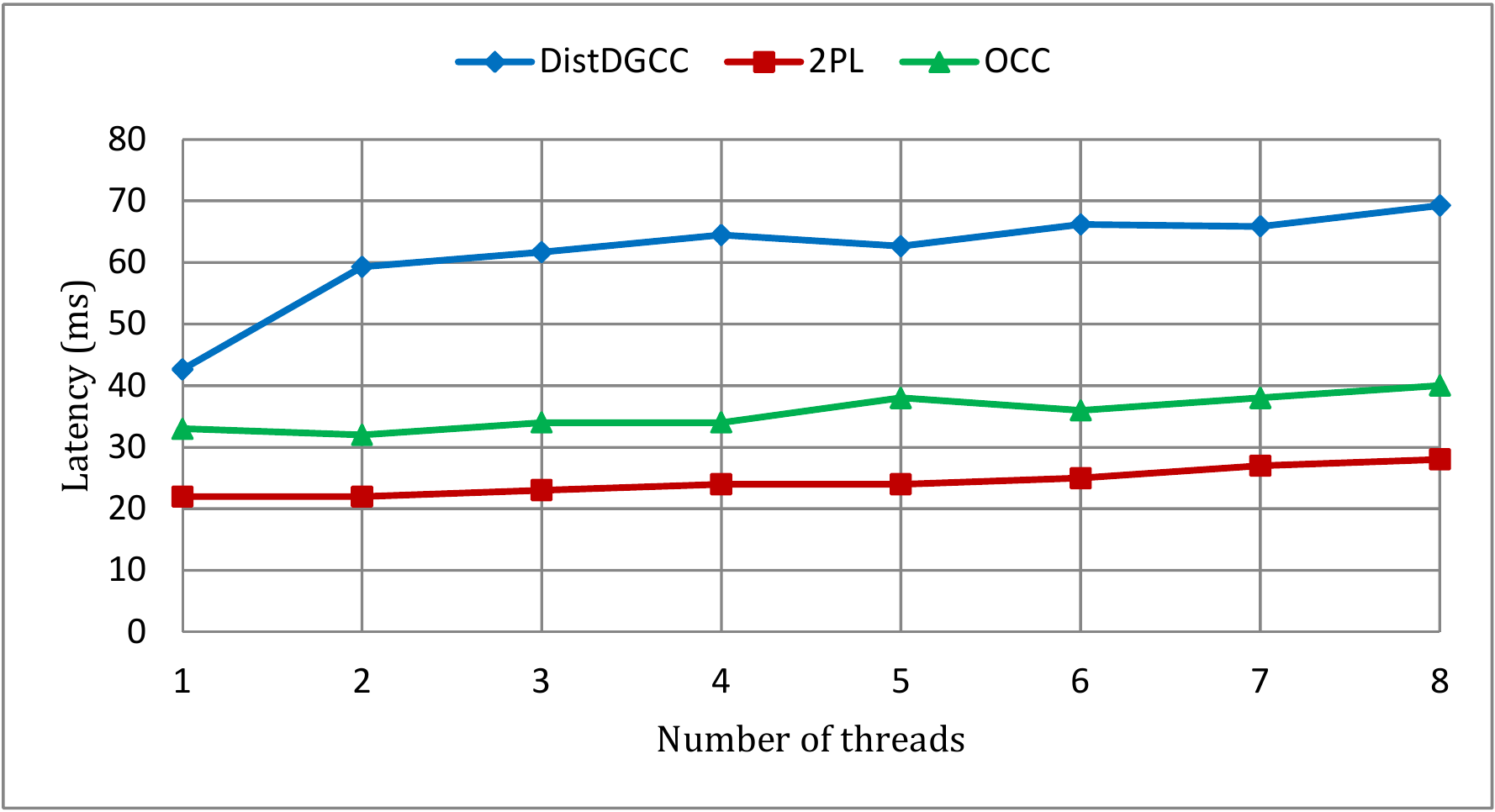}}\quad
	\subfigure[YCSB workload with high contention]{
		\label{fig:ycsb_local_8_lat} 
		\includegraphics[scale=0.3]{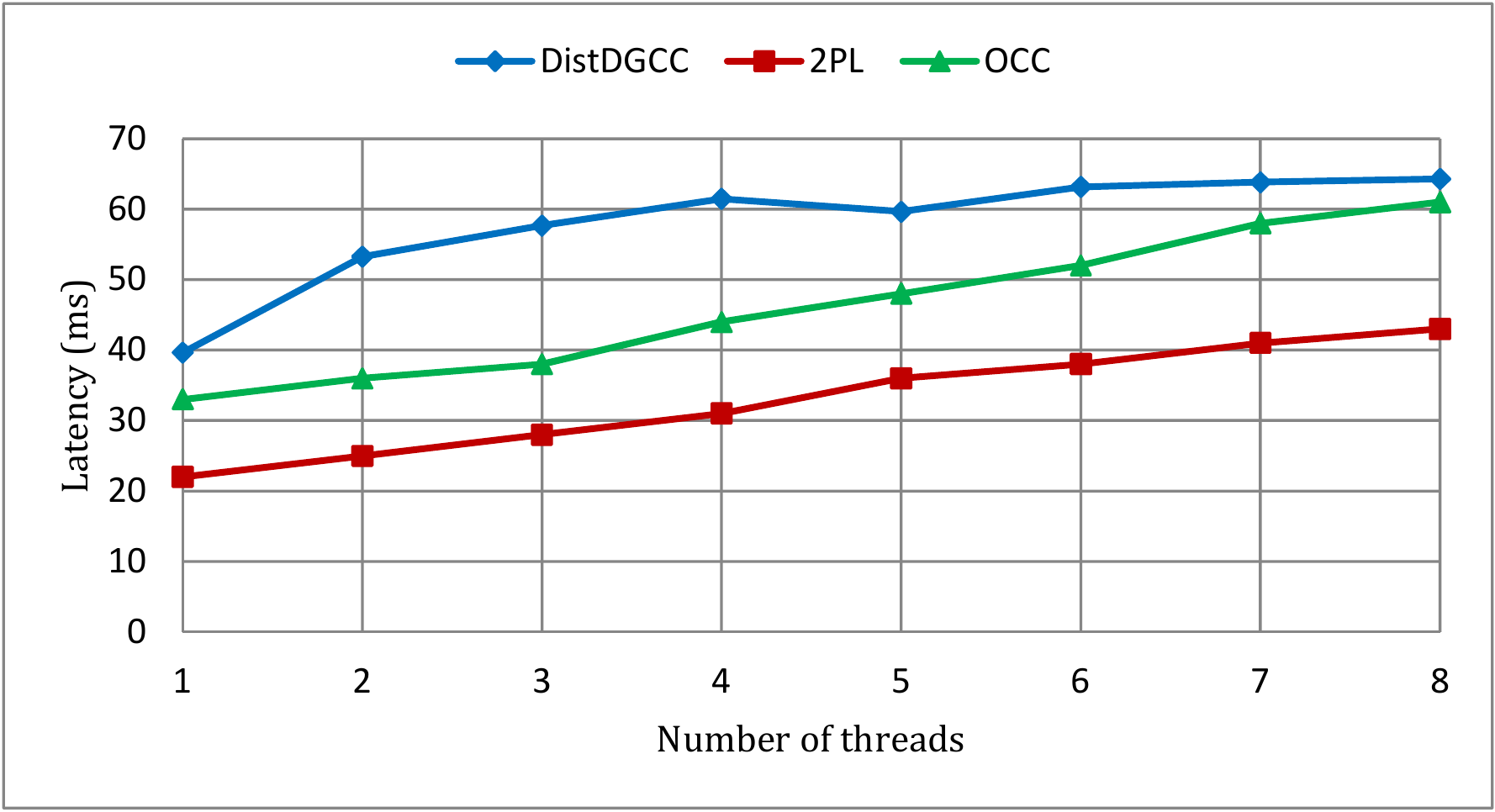}}\quad
	\subfigure[TPC-C workload]{
		\label{fig:tpcc_local_lat} 
		\includegraphics[scale=0.3]{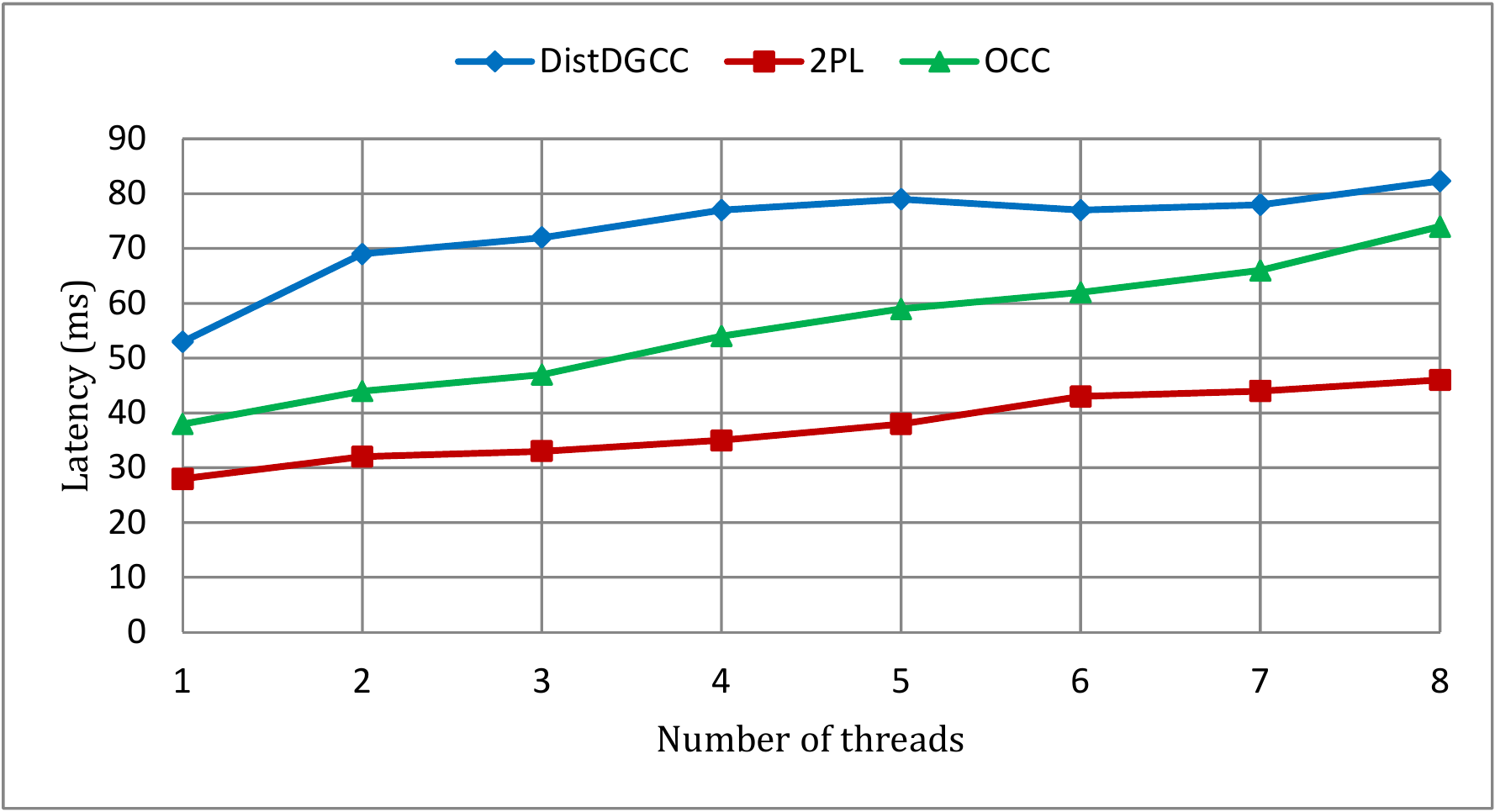}}
	\vspace{-1mm}
	\caption{Latency evaluation}
	\label{fig:distdgcc_local_lat}
\end{figure*}

\subsection{Recovery of Cascading Failure}

In distributed environment, it is common for cascading failure to occur.
There are two different cases for cascading failure in a cluster.
One is that new failure happens on the failed node before the completion of its recovery.
In this case, if there are no newly arrived transactions, the system can handle such cascading failure by simply restarting the recovery on the failed node.
Otherwise, the next recovery should also consider those new generated log records.
The second case is that system failure happens on another node during the recovery.
Although dependency logging is a form of logical logging approach, 
the dependency relations between different nodes are already resolved by saving data images in remote logging records, and hence, it facilitates independent recovery of failed nodes;  cascading failure is therefore also handled as a result.


\section{Evaluation}
\label{sec:eval}

In this section, we first evaluate the performance of \distdgcc\ against 2PL and OCC.
Then we evaluate the performance of dependency logging in both runtime and recovery phase.

\subsection{Experimental Setup}

\noindent{\textbf{Testbed:} We run \distdgcc\ and all logging experiments with our system which is implemented with 18,151 lines of C++.
To evaluate the performance of \distdgcc, 
we compare it with 2PL and OCC which are originally implemented in an open source DBMS~\cite{vldb2014:yuxiangyao}.
To enable a fair comparison, we modified the codes to use the same storage and network system as our implementation.
The performance of dependency logging is evaluated against \aries\ both in runtime and recovery.
System failure is simulated by killing the daemon process on a node 
and the recovery process is then invoked immediately.
All the experiments are conducted on an in-house cluster of 8 nodes. 
Each node has an Intel(R) Xeon(R) 1.8 GHz 4-core CPU, 8GB RAM and 500GB HHD.
}

\smallskip

\noindent{\textbf{Benchmarks:} We adopt two popular benchmarks, namely YCSB~\cite{YCSB} and TPC-C~\cite{TPCC} to conduct the evaluations.
YCSB transaction performs $10$ mixed read/write operations and the key access follows the Zipfian distribution.
To generate both low contention and high contention workloads,
we set the Zipfian parameter to $0.6$ and $0.8$, respectively.
TPC-C is adopted to simulate a real and complete order-entry environment.
TPC-C workload mixes five kinds of transactions: New-Order (44\%), Payment (45\%), Delivery (4\%), Order-Status (4\%) and Stock-Level (3\%). 
}

%

\begin{figure}[t]
	\begin{minipage}{0.47\linewidth}
		\subfigure[Throughput evaluation]{
			\label{fig:ycsb_dist} 
			\includegraphics[width=0.95\linewidth,height=0.75\linewidth]{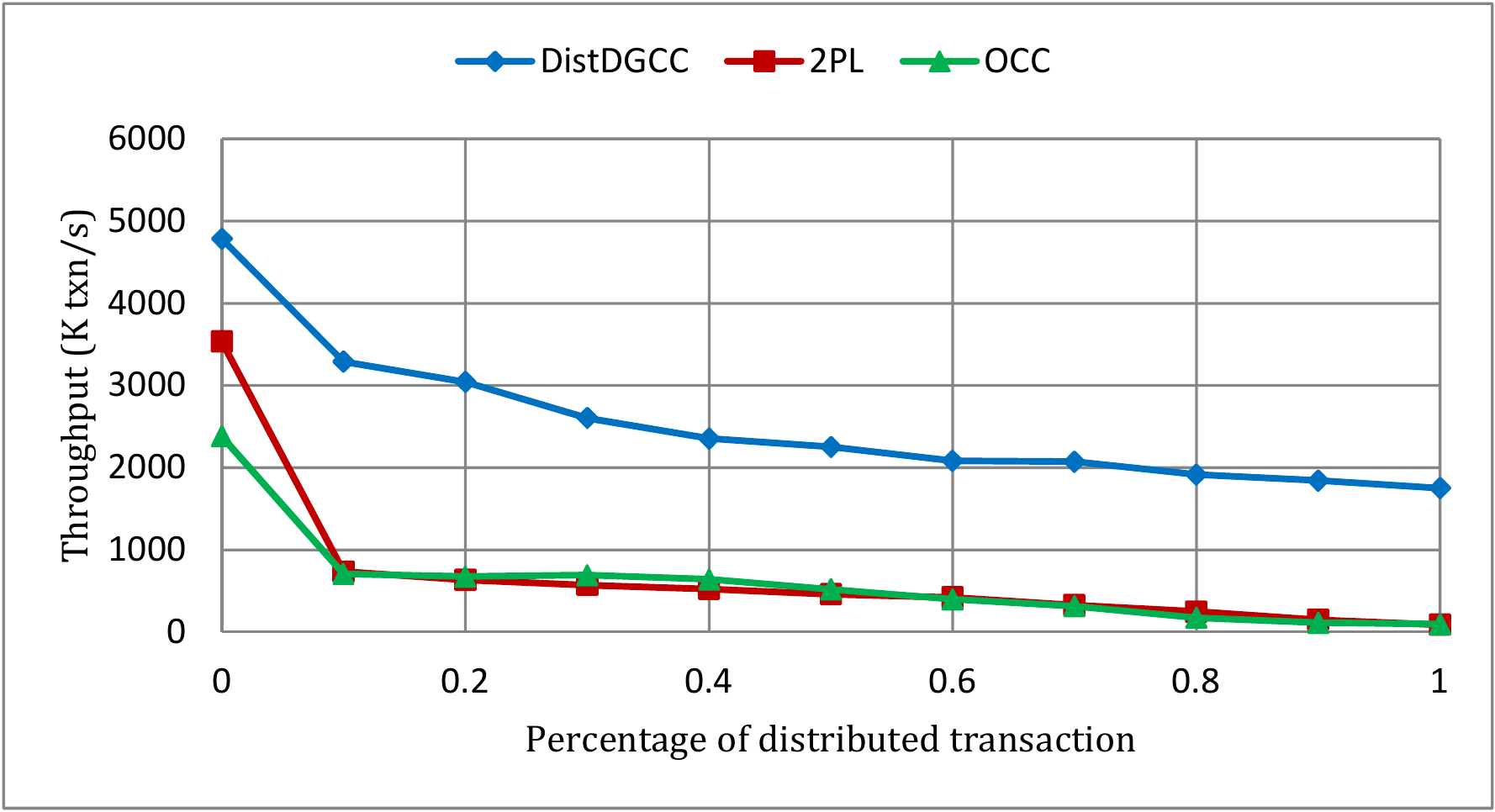}}
		\subfigure[Latency evaluation]{
			\label{fig:ycsb_dist_lat} 
			\includegraphics[width=0.95\linewidth,height=0.75\linewidth]{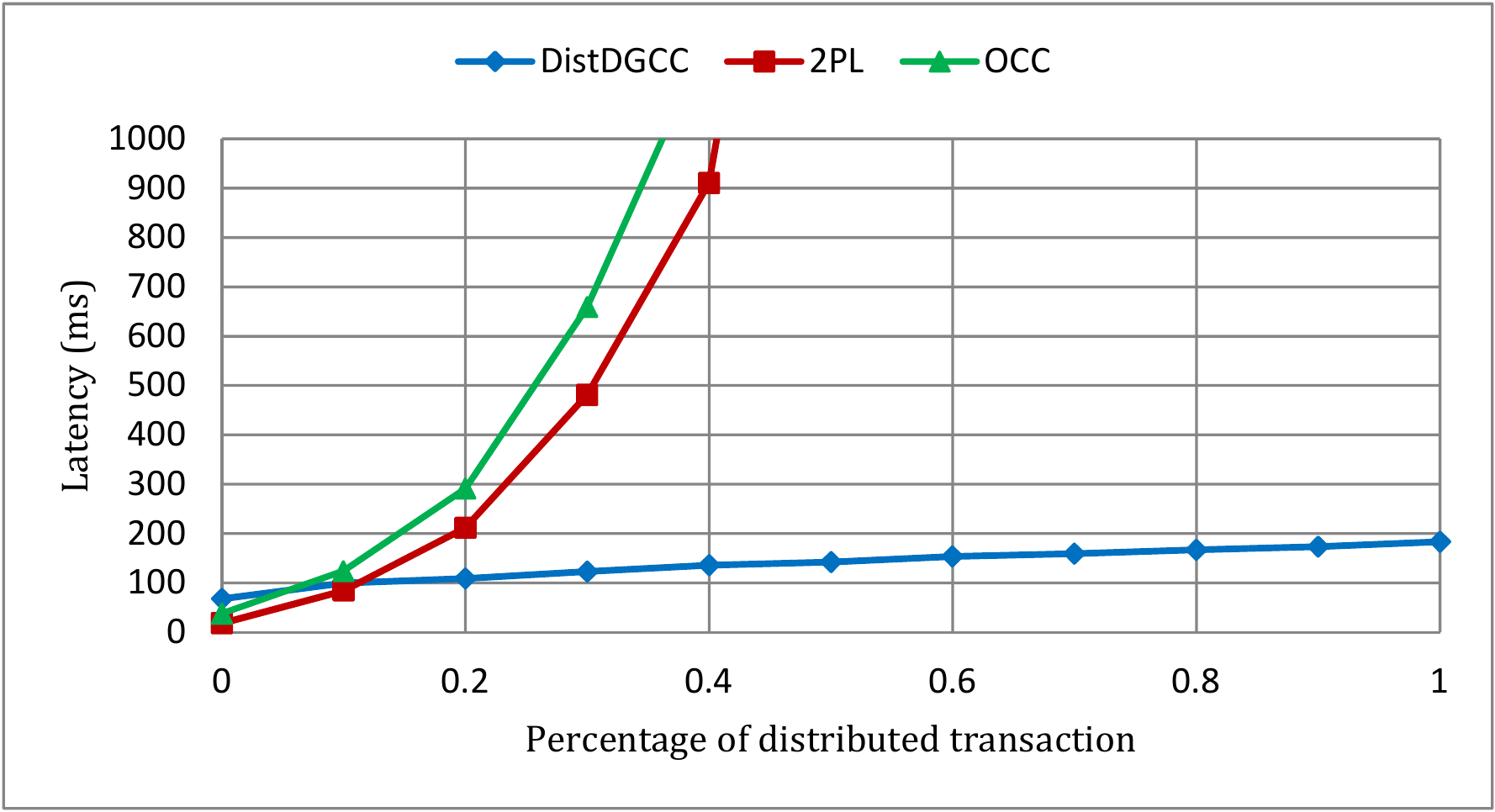}}
		\caption{YCSB workloads on 8 nodes cluster}
		\label{fig:ycsb_dist_all}
	\end{minipage}
	\qquad
	\begin{minipage}{0.47\linewidth}
		\subfigure[Throughput evaluation]{
			\label{fig:tpcc_dist} 
			\includegraphics[width=0.95\linewidth,height=0.75\linewidth]{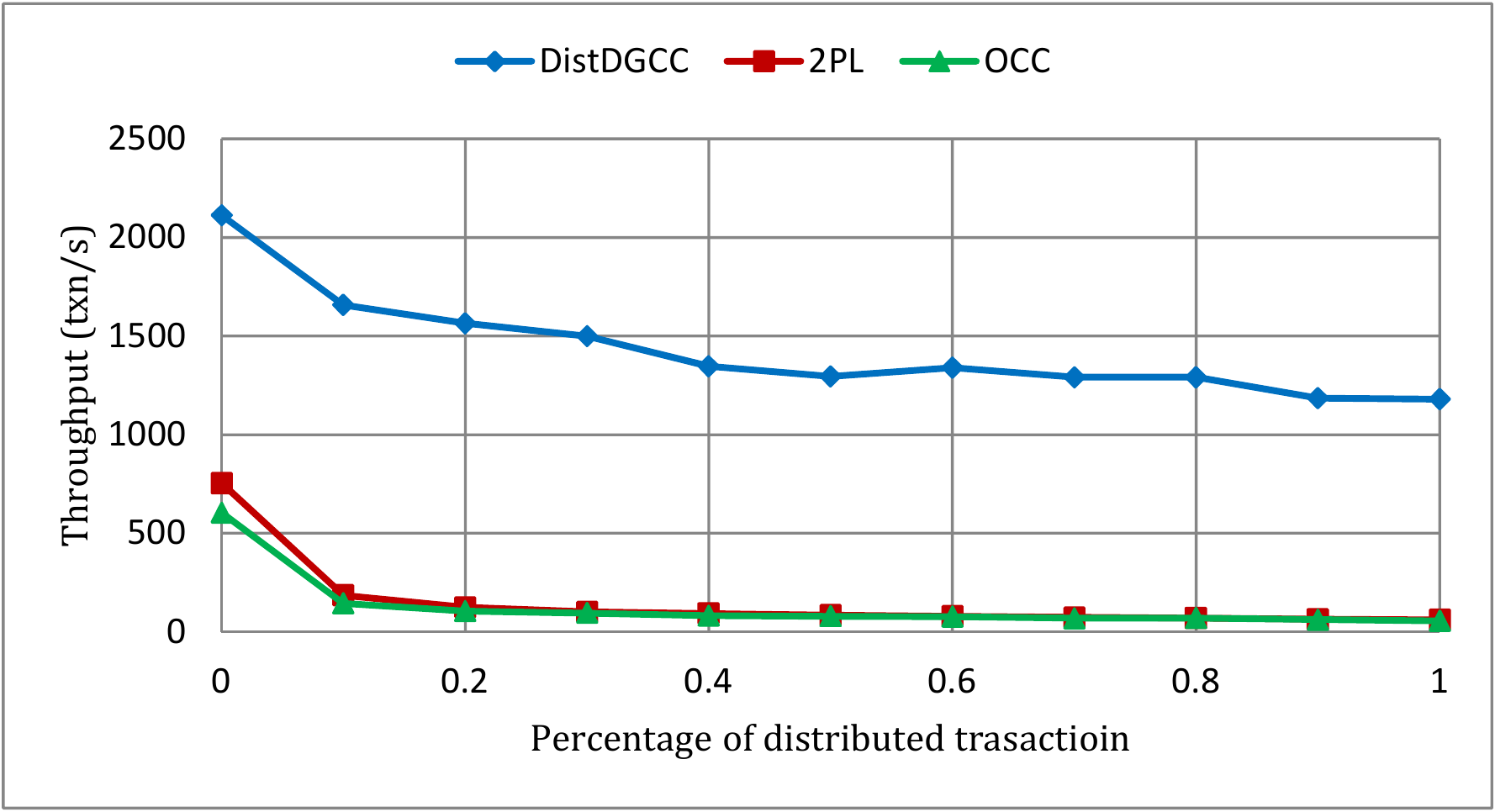}}
		\subfigure[Latency evaluation] {
			\label{fig:tpcc_dist_lat}
			\includegraphics[width=0.95\linewidth,height=0.75\linewidth]{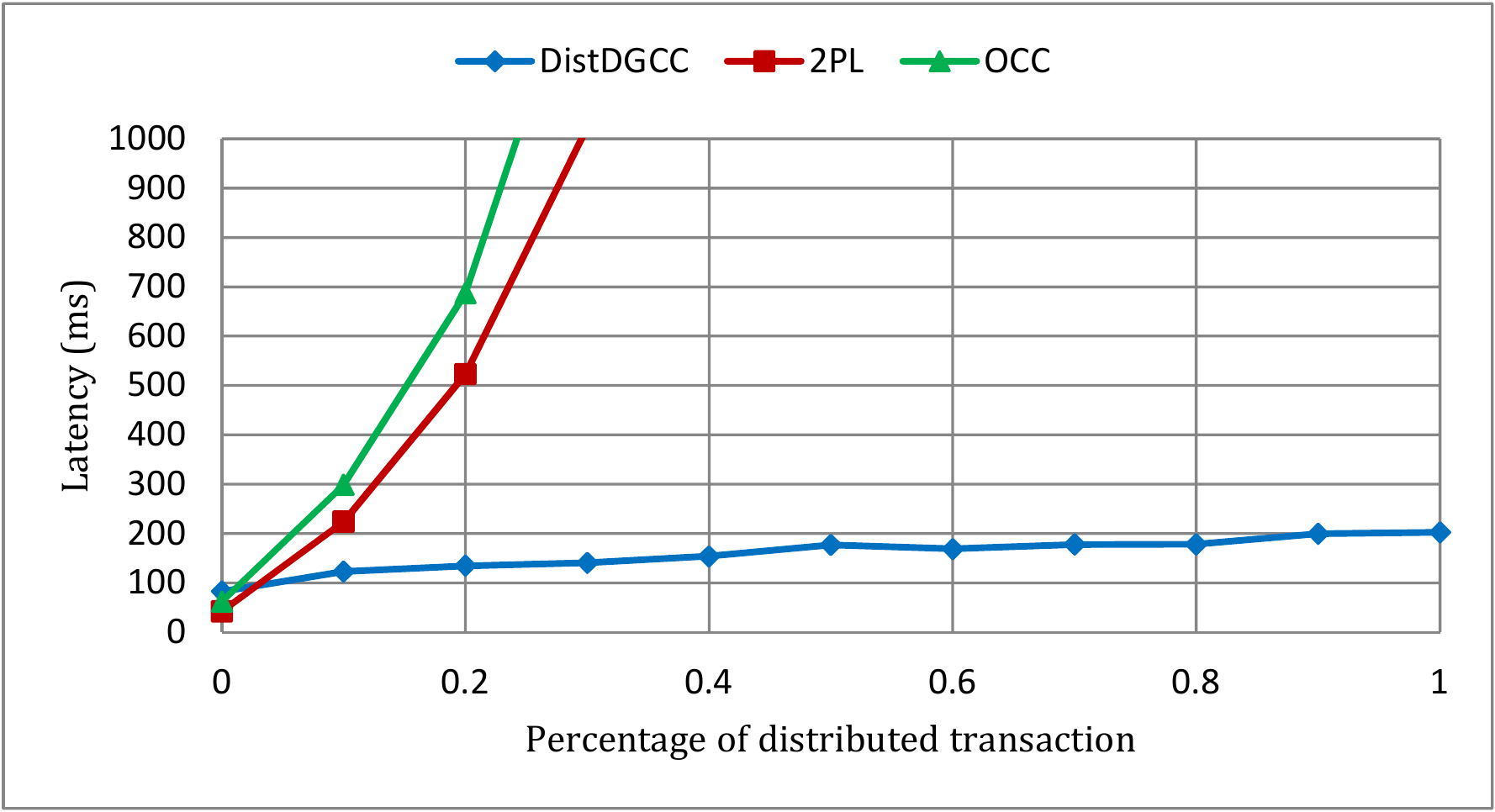}}
		\caption{TPC-C workloads on 8 nodes cluster}
		\label{fig:tpcc_dist_all}
	\end{minipage}
	\vspace{-2mm}
\end{figure}

\subsection{\distdgcc\ Evaluation}
In this section, we evaluate the performance of \distdgcc\ against 2PL and OCC on both single node and cluster.

\smallskip
\noindent\textbf{Evaluation on a single node:}
Figure~\ref{fig:ycsb_local_5} and Figure~\ref{fig:ycsb_local_8} show the throughputs of the three protocols on a single node
with YCSB workload.
In summary, \distdgcc\ shows the best performance in both low contention and high contention settings.
The performance gain mainly comes from the separation of contention resolution and transaction execution, 
which reduces the time on thread blocking.
Moreover,
the acyclicity of the dependency graph eliminates transaction aborts due to contentions 
and further improves the computation efficiency.
2PL also shows good scalability in the low contention setting.
However, 
its performance drops with increasing contentions due to the higher cost of lock acquisition and deadlock resolution.
Compared to \distdgcc\ and 2PL,
OCC performs the worst, since it resolves conflicts during a validation phase using timestamps,
which are usually assigned by a centralized component and may result in a performance bottleneck.
Moreover,
the aborted transactions waste the computation resources and also incur higher penalty.
Figure~\ref{fig:tpcc_local} shows the results with TPC-C workload that contains $1$ warehouse.
In this scenario, the contention rate is high, 
since all the NewOrder and Payment transactions need to  update the data record in the warehouse table.
Hence, \distdgcc\ also exhibits superiority over 2PL and OCC.


In Figure~\ref{fig:distdgcc_local_lat}, we compare the latency of the three protocols running on a single node.
When the contention rate is low,
the latency of the three methods varies within a small range.
However, the latency of 2PL and OCC increases with increasing contentions, because 2PL needs to spend more time on acquiring the locks, while high contention workload leads to more transaction aborts in OCC.
In contrast,
\distdgcc shows its robustness with respect to latency, as it is mainly affected by the batch size.


\smallskip
\noindent\textbf{Evaluation on a cluster:}
Figure~\ref{fig:ycsb_dist_all} and Figure~\ref{fig:tpcc_dist_all} show the performances of the three protocols on an 8-node cluster by varying the percentage of distributed transaction.
As shown in Figure~\ref{fig:ycsb_dist} and Figure~\ref{fig:tpcc_dist},
the throughput of 2PL and OCC are affected significantly by distributed transactions.
Even a small portion of distributed transactions leads to a dramatical performance loss.
Compared to a local transaction,
a distributed transaction accesses data records in multiple nodes
and incurs extra network cost.
More importantly, the worker thread has to be blocked until the distributed transaction commits or aborts,
which degrades the computation utilization and hence leads to the performance loss.
Since
 \distdgcc\ processes transactions in a batch manner, network messages for one batch of transactions are aggregated, 
which leads to an improved network performance.
Moreover, worker threads will not be blocked during the processing.
Instead, they construct the dependency graphs for the next batch of transactions, and the computation utilization is improved as a result.
Hence, \distdgcc\ exhibits good performance superiority  over 2PL and OCC in the distributed setting. 
While \distdgcc\ shows good robustness to the percentage of distributed transactions, 
its performance also drops as more distributed transactions are involved due to the fact that more network messages are generated.

In Figure~\ref{fig:ycsb_dist_lat} and Figure~\ref{fig:tpcc_dist_lat},
we show the latency variations.
The latency of both 2PL and OCC increases significantly when there are more distributed transactions.
The reason are twofold.
First, a distributed transaction is more expensive and always incurs higher latency.
Second, a worker thread has to wait until the distributed transaction commits or aborts and thus increases the latency for those blocked transactions.
\distdgcc\ reduces the network overheads by aggregating network messages
and also avoids long waiting time caused by thread blocking.
Consequently, its latency increases only slightly with the increasing percentage of distributed transactions.

\begin{figure}[t]
	\centering
    \subfigure[Throughput on YCSB]{
       \centering
       \label{fig:ycsb_batch} 
       \includegraphics[width=0.47\linewidth,height=0.35\linewidth]{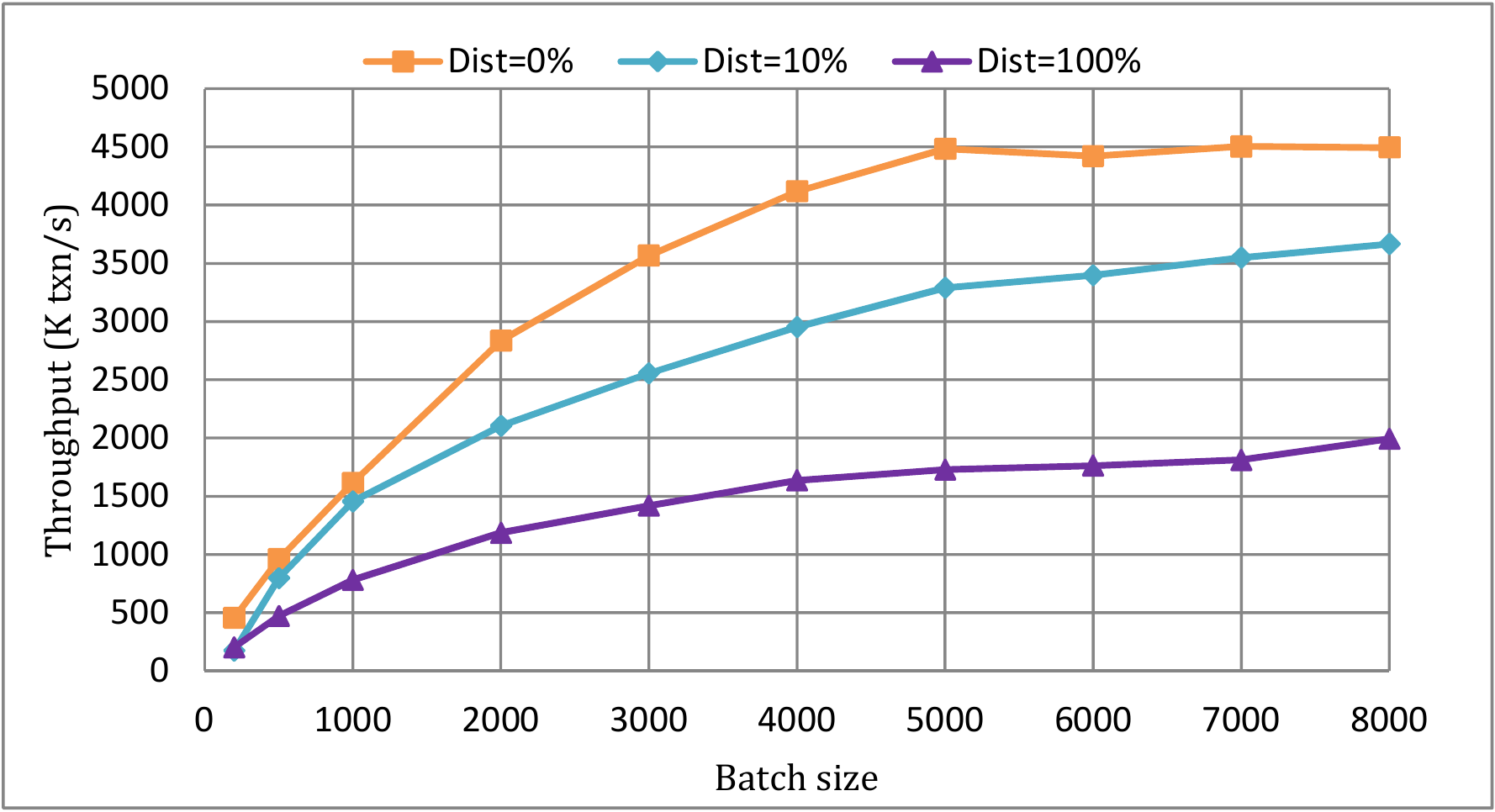}}\quad
     \subfigure[Latency on YCSB]{
     	\centering
       \label{fig:ycsb_batch_lat} 
       \includegraphics[width=0.47\linewidth,height=0.35\linewidth]{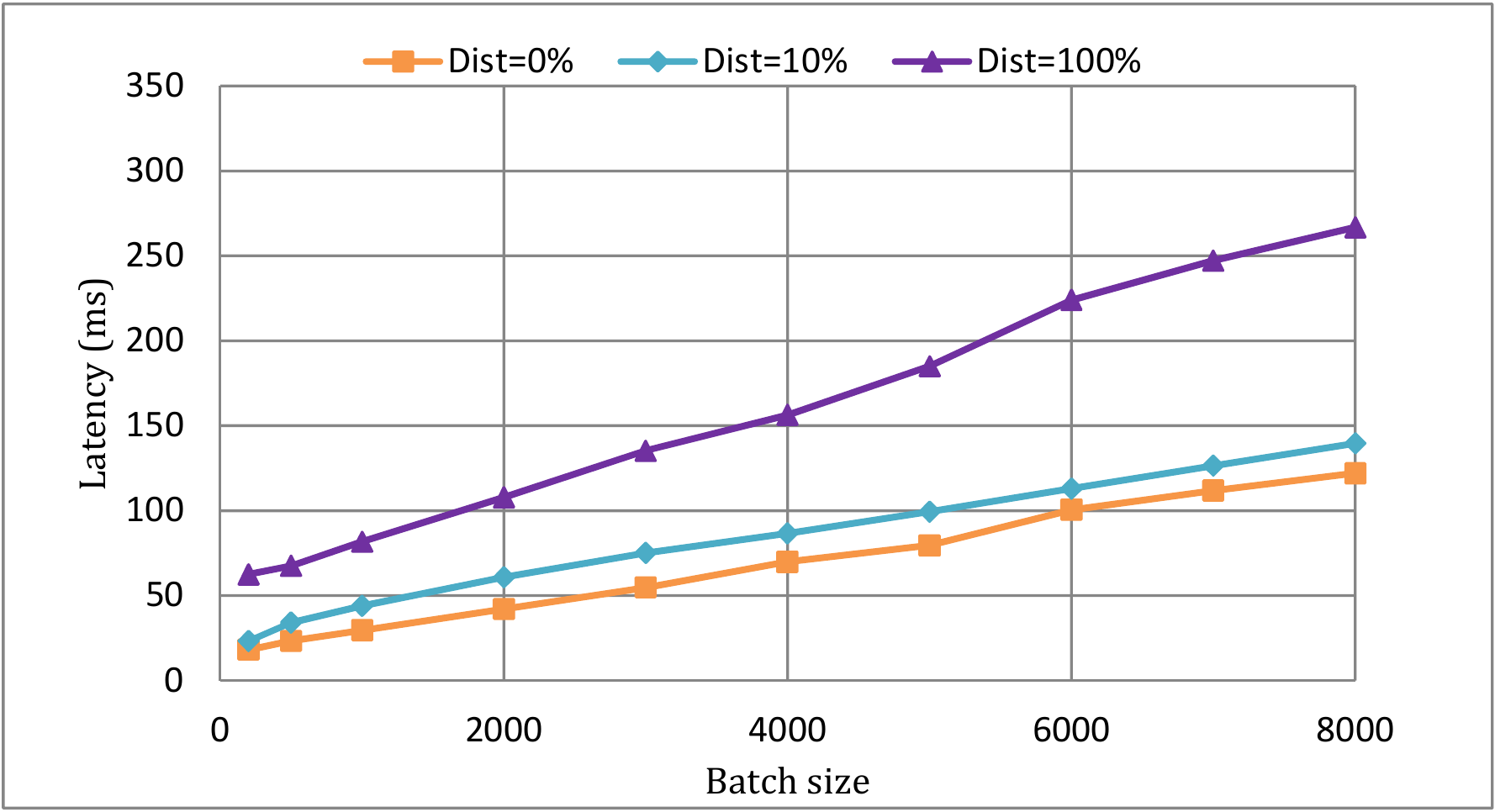}}
       \\
     \subfigure[Throughput on TPC-C]{
     	\centering
        \label{fig:tpcc_batch} 
        \includegraphics[width=0.47\linewidth,height=0.35\linewidth]{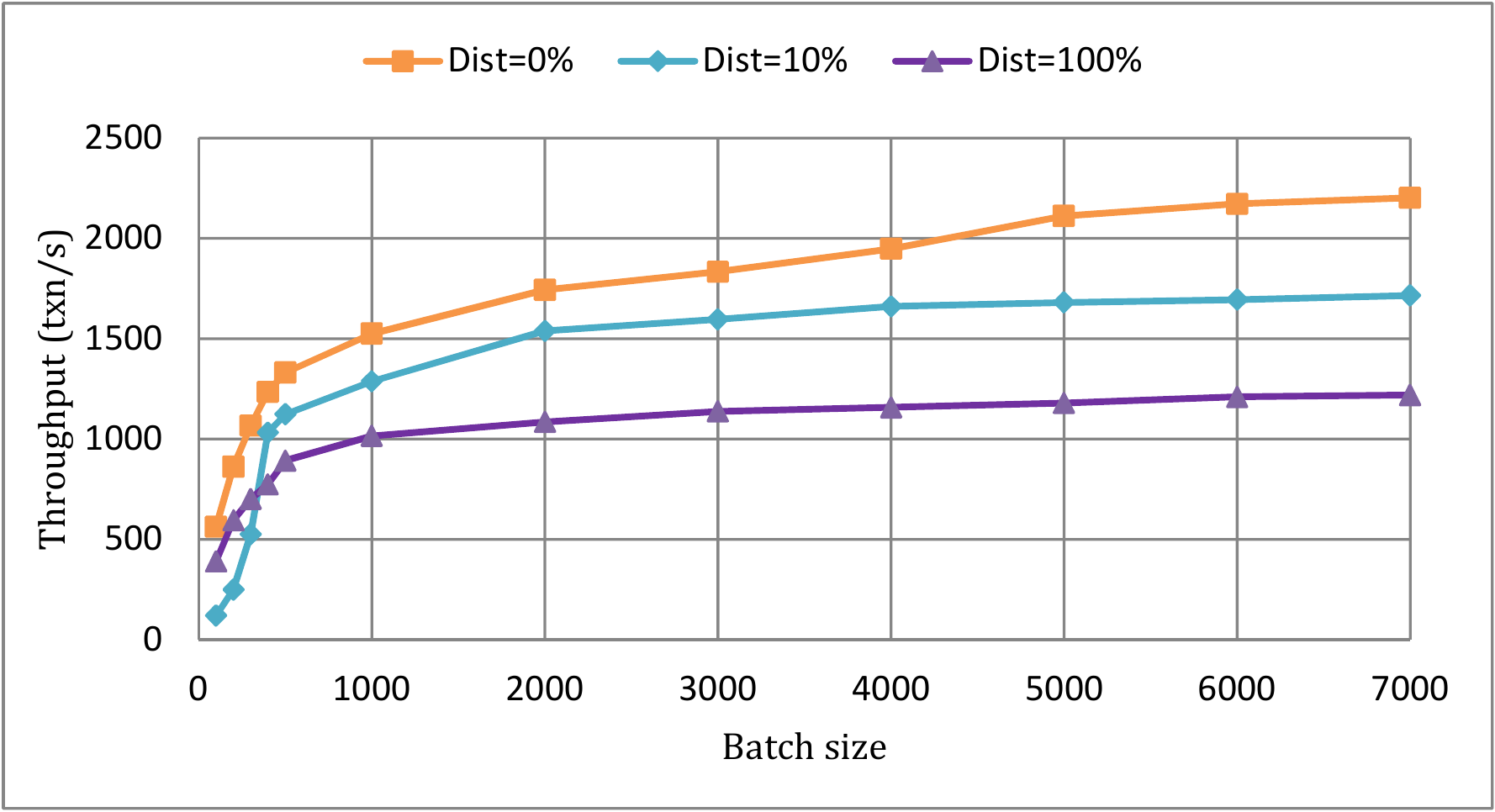}}\quad
     \subfigure[Latency on TPC-C] {
     	\centering
        \label{fig:tpcc_batch_lat}
        \includegraphics[width=0.47\linewidth,height=0.35\linewidth]{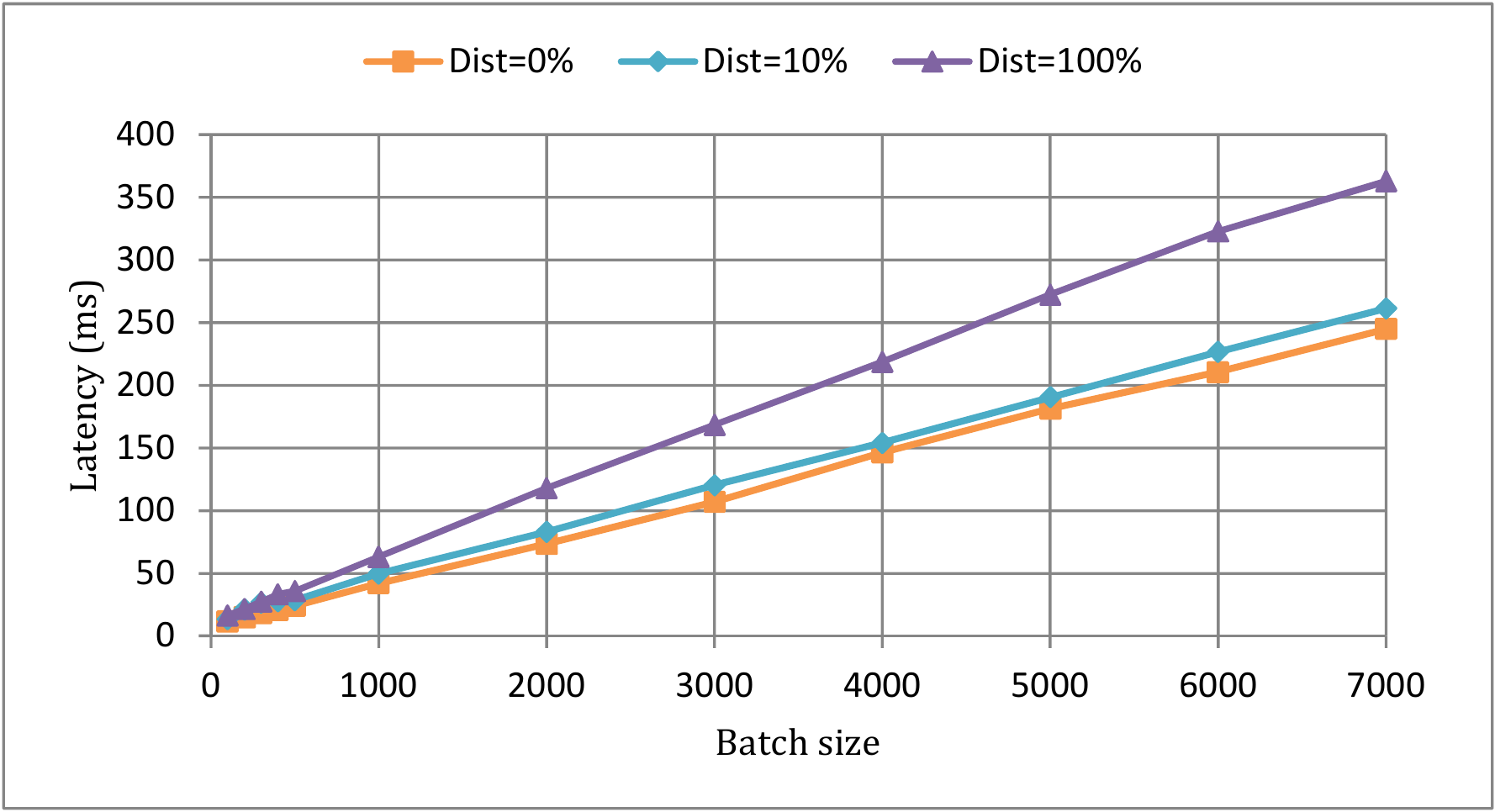}}
    	\vspace{-4mm}
     \caption{Effects of batch size on an 8 nodes cluster}
     \label{fig:batch}
\end{figure}

\smallskip
\noindent\textbf{Effects of batch size:}
The effects of batch size on the throughput and latency for \distdgcc\ are shown in Figure~\ref{fig:batch} .
We fix the number of threads on each node to $8$.
As shown in Figure~\ref{fig:ycsb_batch} and Figure~\ref{fig:tpcc_batch}, 
the throughput of \distdgcc\ first increases with the batch size due to better exploitation of computation resources.
It subsequently becomes plateauing since computation resources on each node are limited.
For the same reason, as illustrated in Figure~\ref{fig:ycsb_batch_lat} and Figure~\ref{fig:tpcc_batch_lat},
the latency increases almost linearly with the batch size.

\begin{figure*}[t]
    \subfigure[Throughput on YCSB]{
       \label{fig:ycsb_local_logging} 
       \includegraphics[width=0.23\linewidth,height=0.18\linewidth]{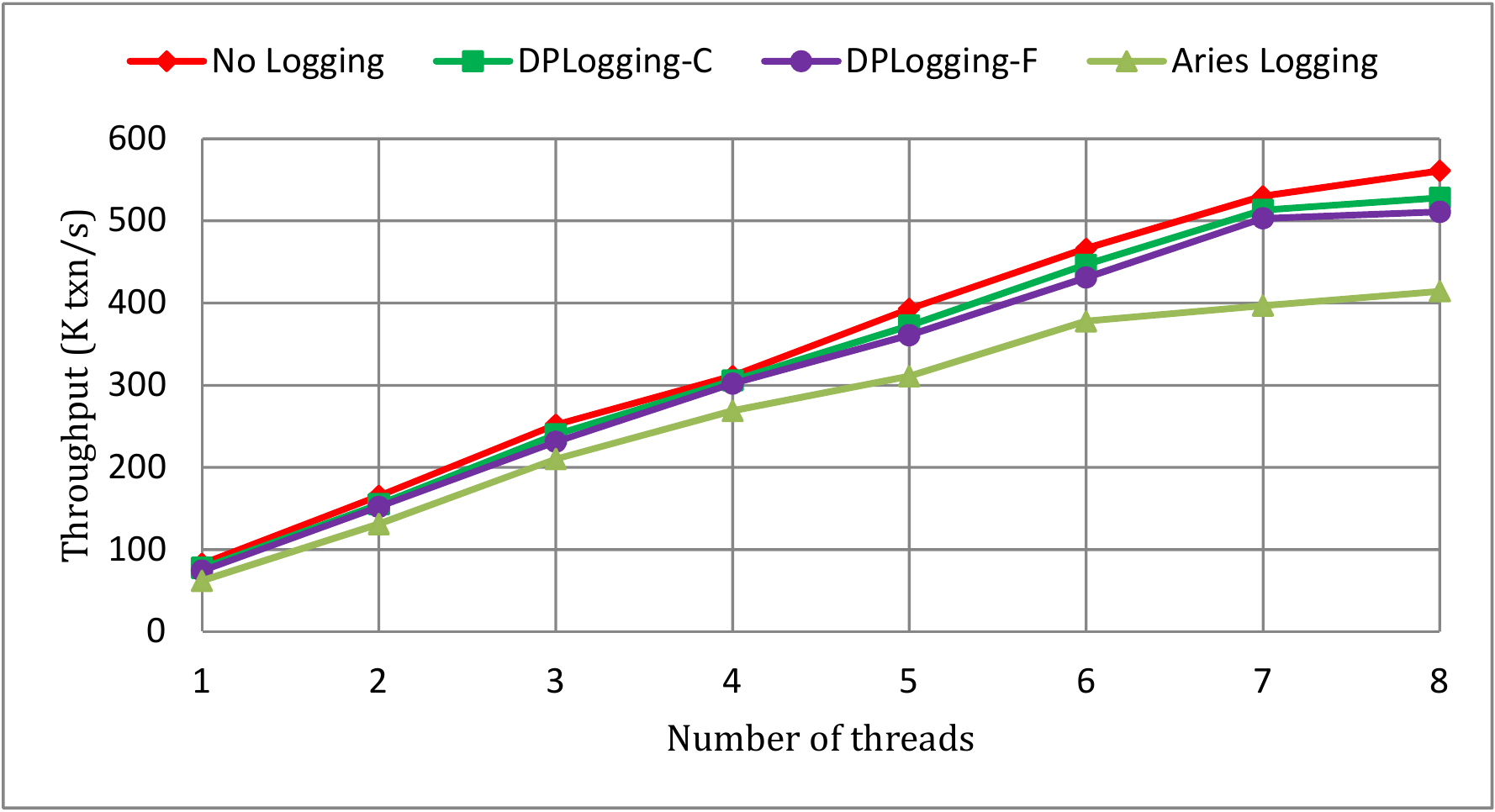}}\quad
     \subfigure[Latency on YCSB]{
       \label{fig:ycsb_local_logging_lat} 
       \includegraphics[width=0.23\linewidth,height=0.18\linewidth]{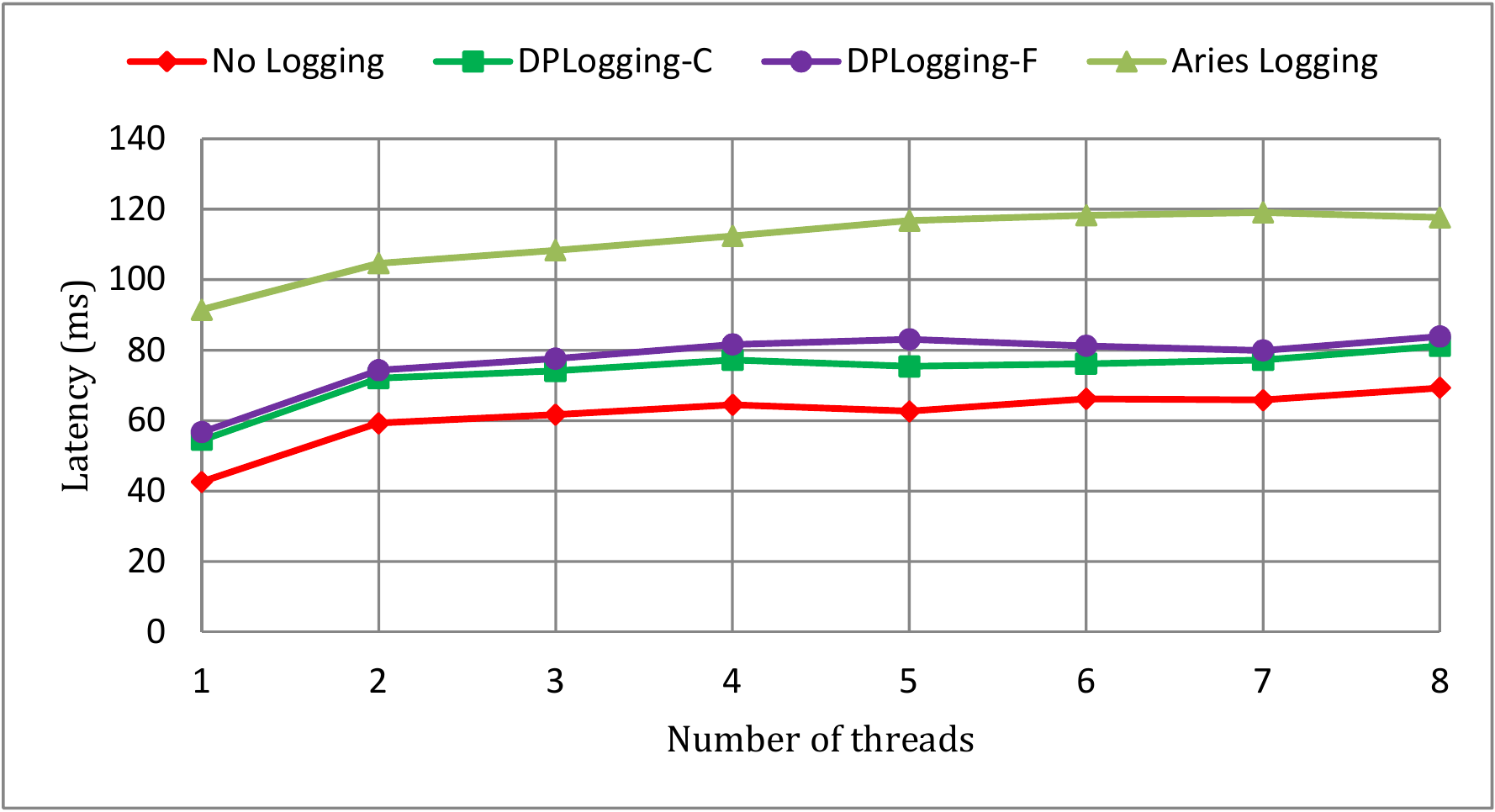}}\quad
     \subfigure[Throughput on TPC-C]{
        \label{fig:tpcc_local_logging} 
        \includegraphics[width=0.23\linewidth,height=0.18\linewidth]{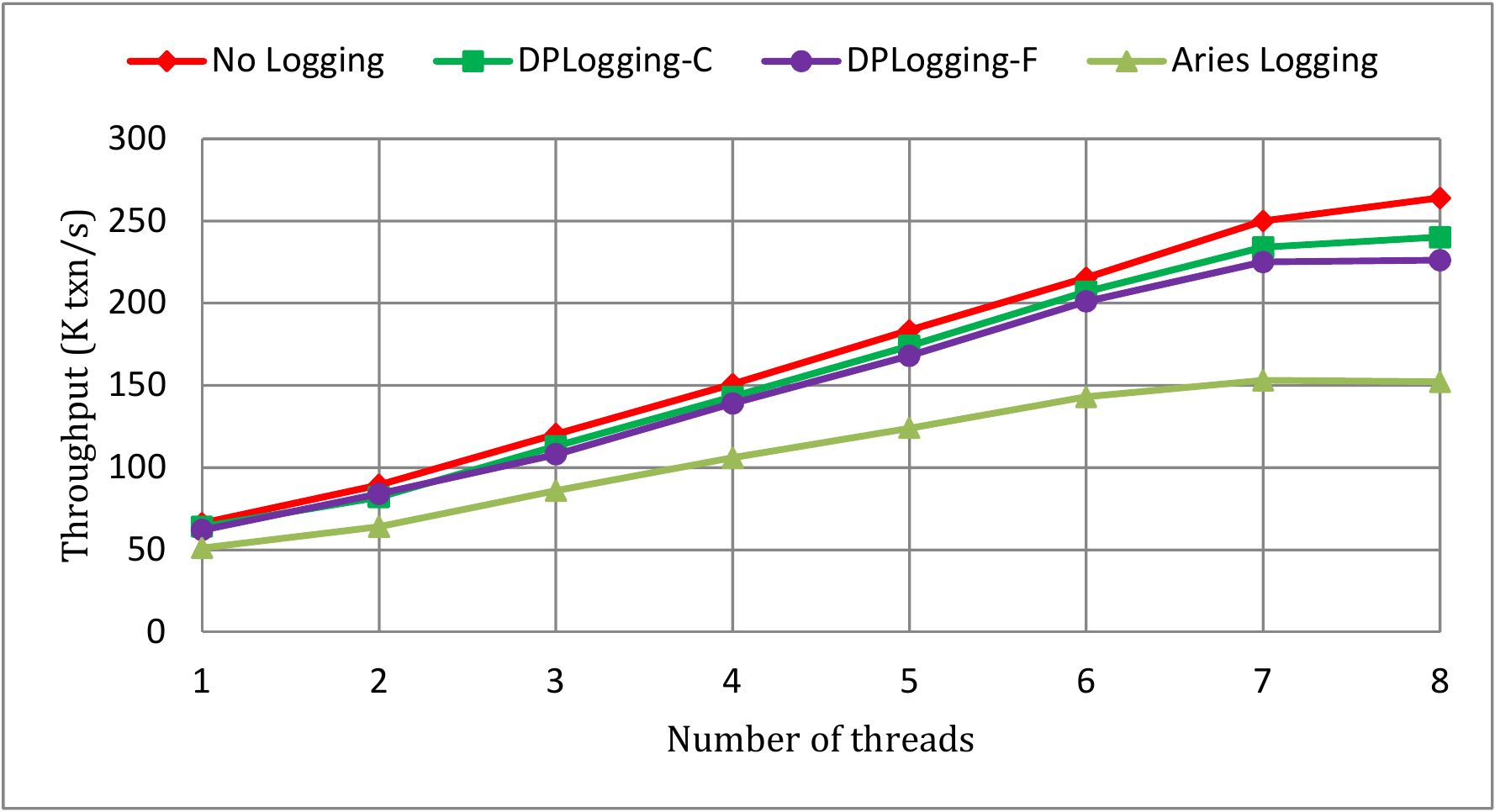}}\quad
     \subfigure[Latency on TPC-C] {
        \label{fig:tpcc_local_logging_lat}
        \includegraphics[width=0.23\linewidth,height=0.18\linewidth]{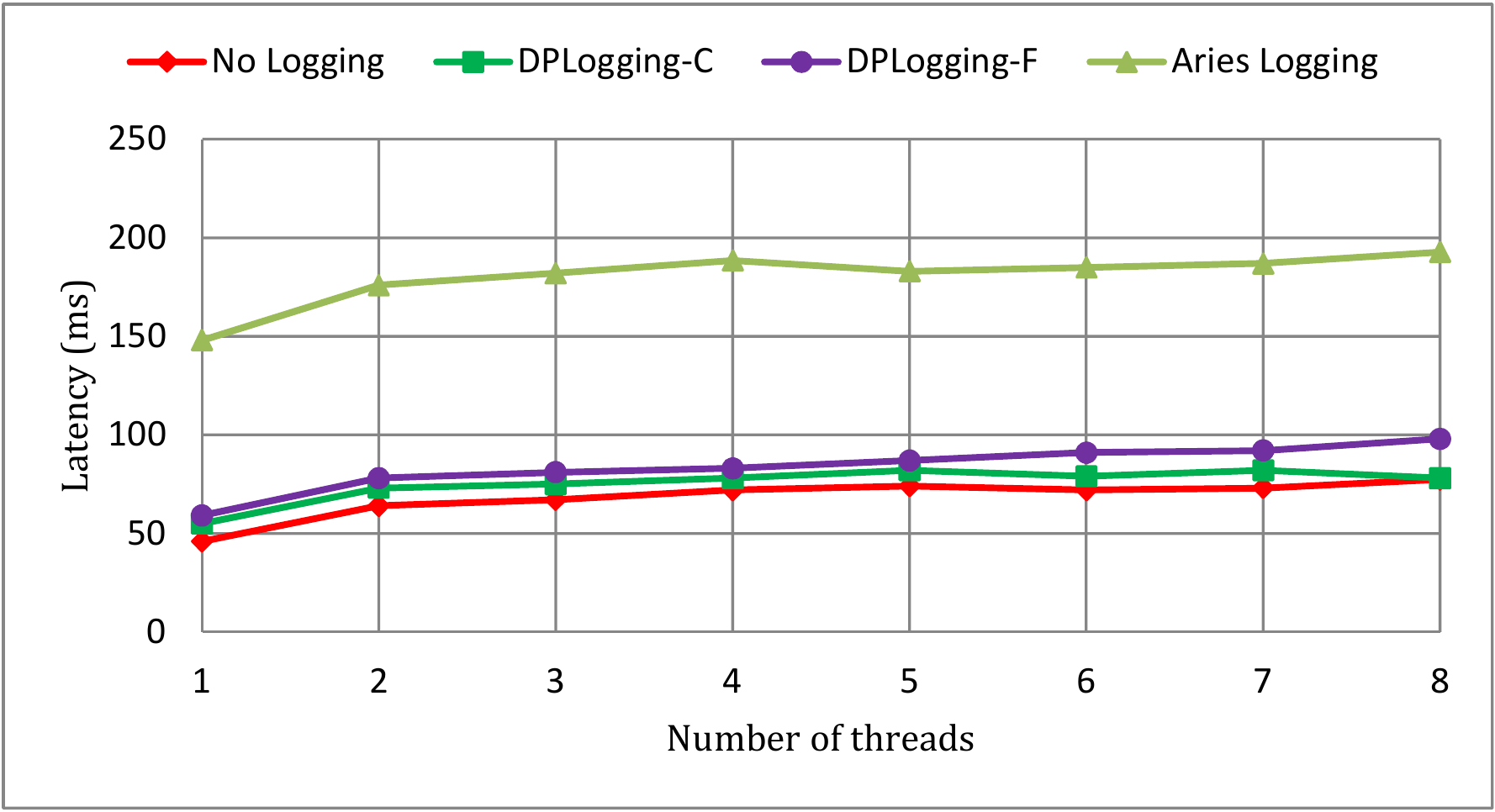}}
    \vspace{-2mm}
     \caption{Effects of logging using only local transactions}
     \label{fig:logging_local}
\end{figure*}

\begin{figure*}[t]
    \subfigure[Throughput on YCSB]{
       \label{fig:ycsb_dist_logging} 
       \includegraphics[width=0.23\linewidth,height=0.18\linewidth]{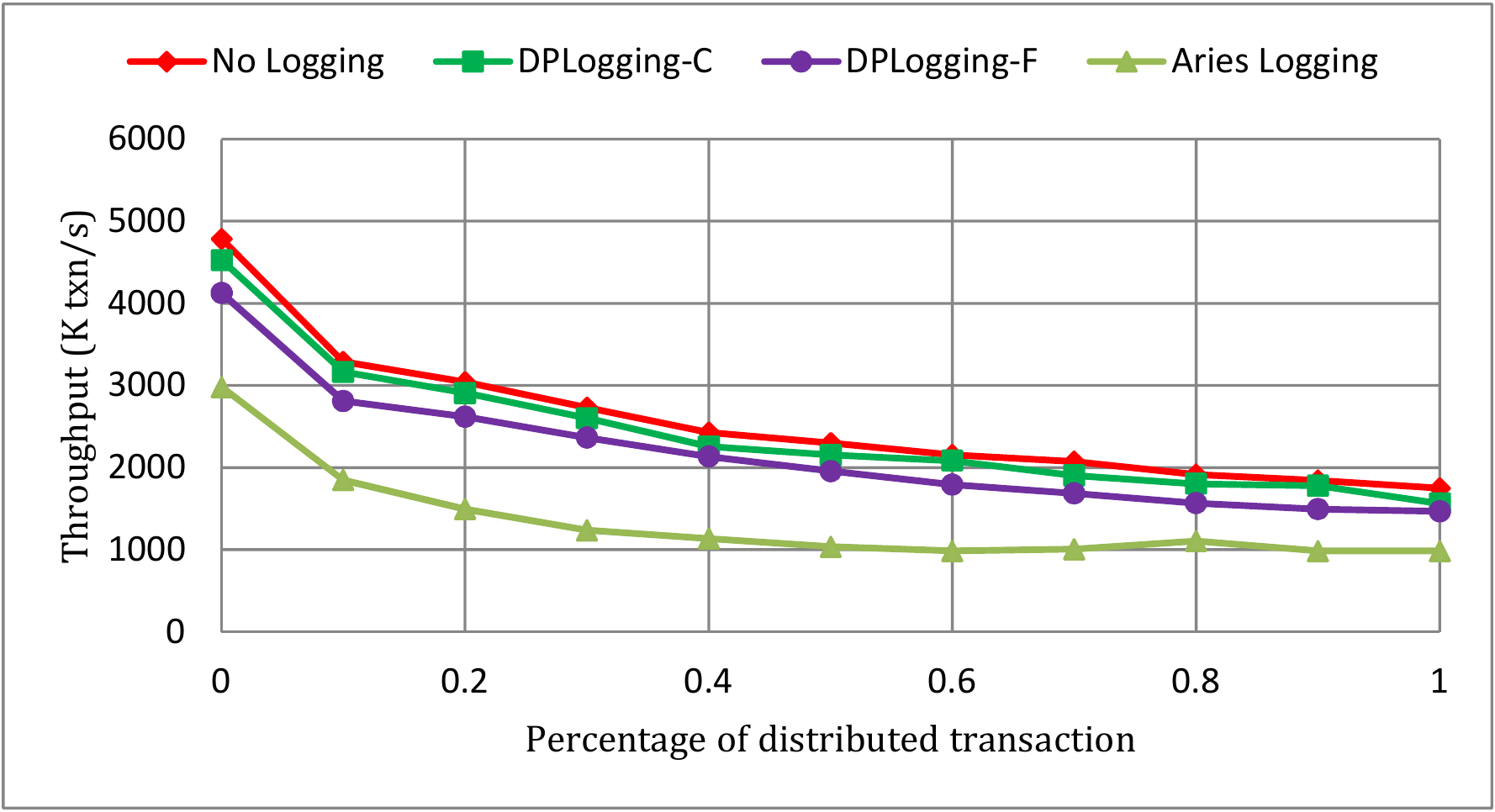}}\quad
     \subfigure[Latency on YCSB]{
       \label{fig:ycsb_dist_logging_lat} 
       \includegraphics[width=0.23\linewidth,height=0.18\linewidth]{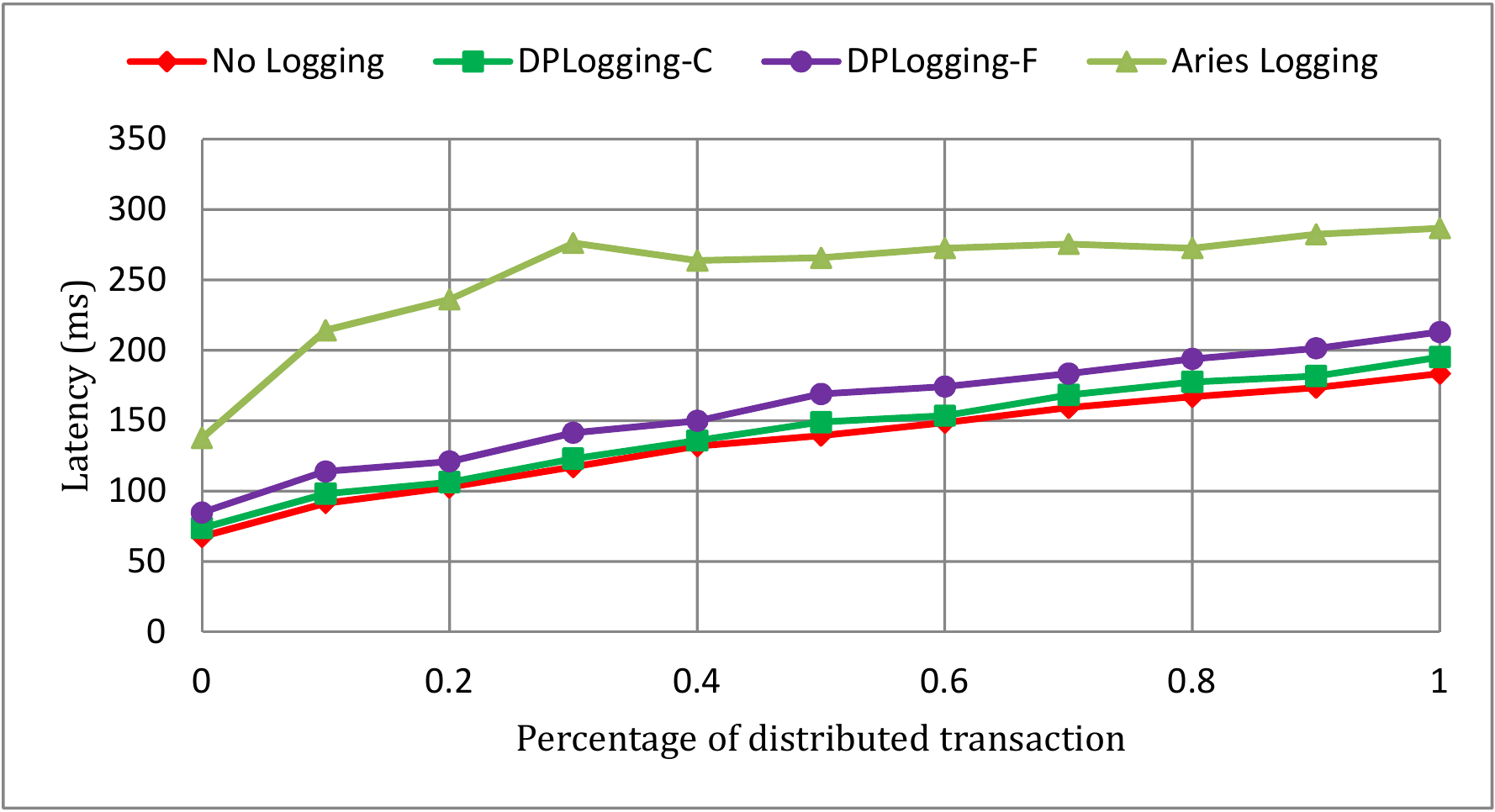}}\quad
     \subfigure[Throughput on TPC-C]{
        \label{fig:tpcc_dist_logging} 
        \includegraphics[width=0.23\linewidth,height=0.18\linewidth]{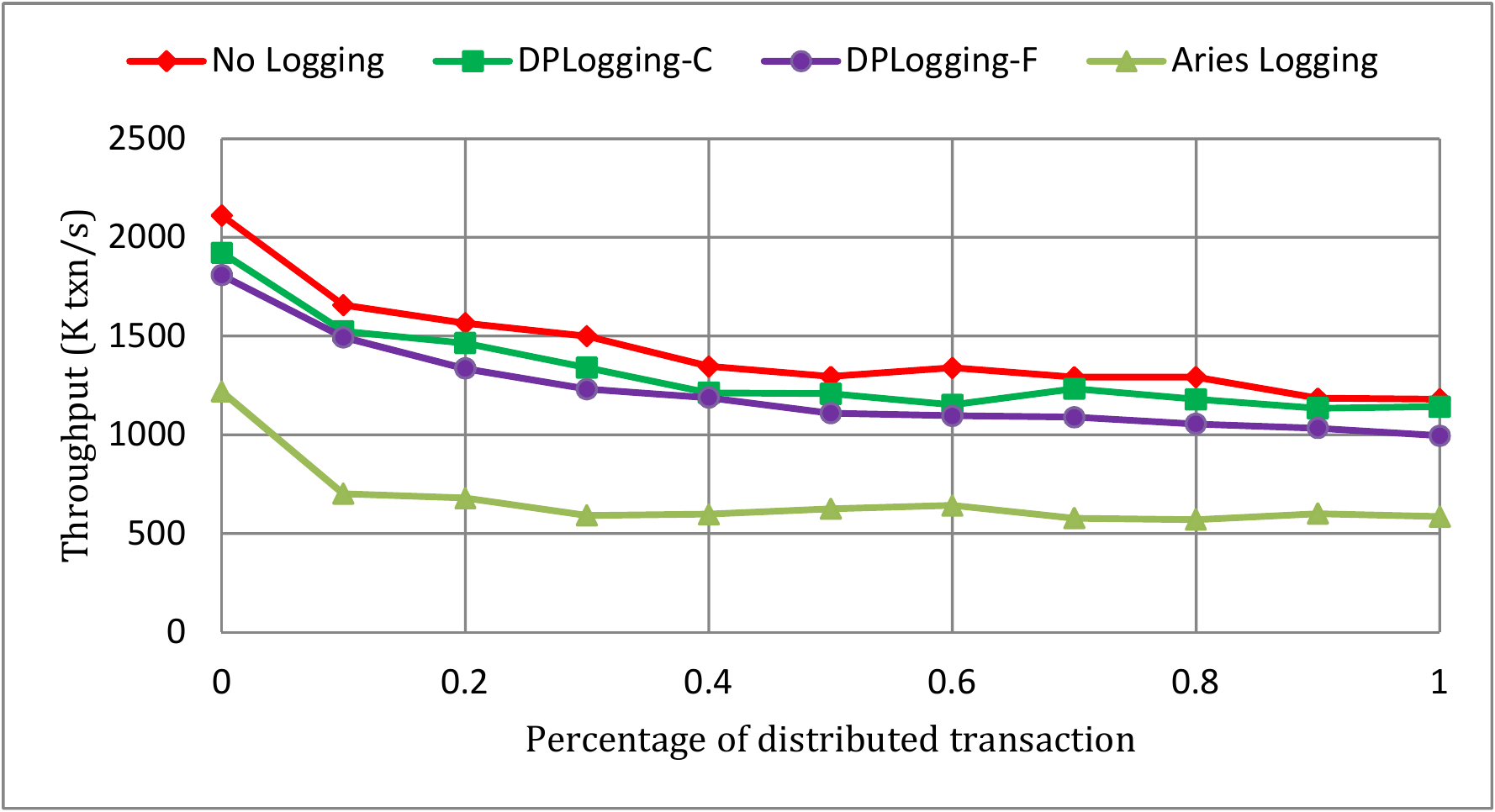}}\quad
     \subfigure[Latency on TPC-C] {
        \label{fig:tpcc_dist_logging_lat}
        \includegraphics[width=0.23\linewidth,height=0.18\linewidth]{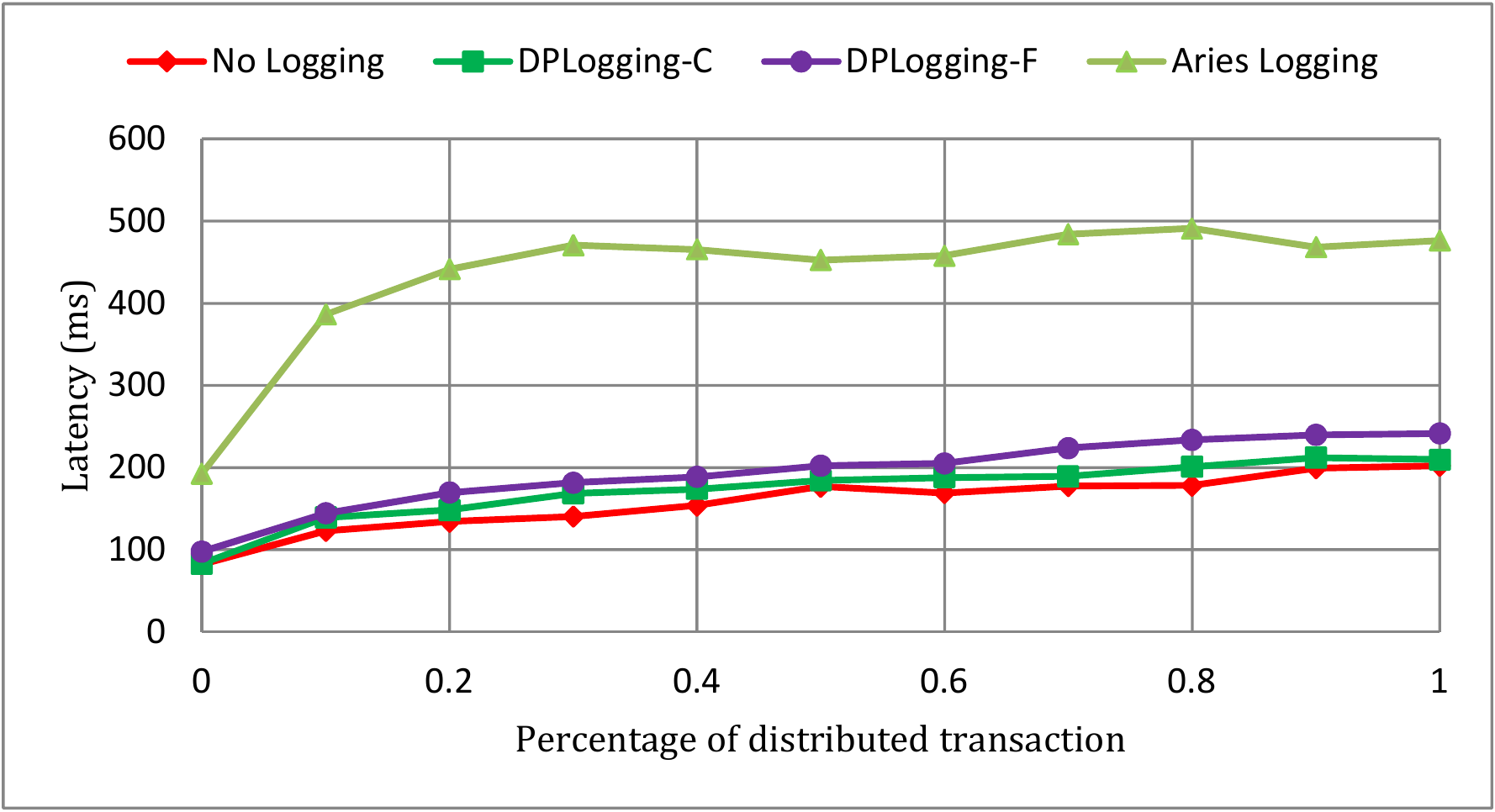}}
     \vspace{-2mm}
     \caption{Effects of logging using mixture of local and distributed transactions}
     \label{fig:logging_dist}
\end{figure*}

\subsection{Dependency Logging Evaluation}

In this subsection, we shall study the efficiency of dependency logging by comparing with \aries.
For ease of explanation, we shall use the following representative names.
\begin{itemize} 
	\item \textbf{No Logging} - disable the logging during the runtime.
	\item \textbf{DPLogging-F} - fine-grained dependency logging approach proposed in Section~\ref{subsec:dplogging-f}.
	\item \textbf{DPLogging-C} - coarse-grained dependency logging approach proposed in Section~\ref{subsec:dplogging-c}.
	\item \textbf{Aries Logging} - \aries\ approach with memory optimizations described in Section~\ref{subsec:log_compression} and Section~\ref{subsec:batch_optimization}.
\end{itemize}

\noindent\textbf{Runtime Evaluation:}
We shall first evaluate the overhead of the proposed dependency logging approach during the runtime.
Two kinds of workloads are used: one workload only contains local transactions, while the other workload contains both local and distributed transactions.

Figure~\ref{fig:ycsb_local_logging} and Figure~\ref{fig:tpcc_local_logging} show the throughputs of the four logging approaches for the workload that only local transactions are involved.
When the number of worker threads is small,
all these logging approaches achieve similar throughputs,
since disk I/Os caused by logging do not cause the performance bottleneck. 
As more worker threads are adopted, Aries Logging generates more disk I/Os and its throughput grows slower than the other three approaches.
Compared to the ideal case that is represented as the No Logging approach,
dependency logging achieves comparative performances.
Unlike Aries Logging that generates one log record for each updated data record,
DPLogging-C creates one log record for each transaction. 
Moreover, each log record only tracks the transaction information instead of the updated data records.
Thus, DPLogging-C generates much less log data and reduces the number of disk I/Os.
While DPLogging-F adopts a fine-grained logging strategy like the Aries Logging,
it reduces the number of log records using the dependency information during the runtime.
If DPLogging-F saves data image for one update in its log record,
there is no need to generate log records for the previous updates on the same record.

Figure~\ref{fig:ycsb_dist_logging} and Figure~\ref{fig:tpcc_dist_logging} show the results for the workload when distributed transactions are involved.
In this set of experiments,
we fix the number of worker threads on each node to $8$ while we vary the percentage of distributed transaction from $0\%$ to $100\%$, to study the effects of logging and network communication on the system performance.
DPLogging-C and DPLogging-F show comparative performance as the No Logging approach.
The gaps between Aries Logging and the other approaches become narrower as more distributed transactions are involved,
since the extra network cost lightens the effect of logging.
Even when all transactions are distributed transactions,
DPLogging-C and DPLogging-F still achieve 1.4X and 1.8X higher runtime performance than that of Aries Logging 
with YCSB workload and TPC-C workload, respectively.

Figure~\ref{fig:ycsb_local_logging_lat}, Figure~\ref{fig:tpcc_local_logging_lat}, Figure~\ref{fig:ycsb_dist_logging_lat} and Figure~\ref{fig:tpcc_dist_logging_lat} show the effects of logging approaches on latency.
While the average latency of DPLogging-C and DPLogging-F is slightly higher than that of No Logging,
they are much lower than that of Aries Logging in both local and distributed settings.
In pure local transaction processing,
the latency of all approaches changes within a small range.
In the distributed setting,
the latency of DPLogging-C and DPLogging-F increase with the percentage of distributed transactions,
since they are mainly dominated by the network communication cost.
However, the latency of Aries Logging appears fairly stable irrespective of the percentage of distributed transactions, because Aries Logging incurs a lot of disk I/Os that dominates its latency.

\begin{figure}[t]
	\subfigure[Recovery with 1 thread]{
		\label{fig:ycsb_recovery_1} 
		\includegraphics[scale=0.23]{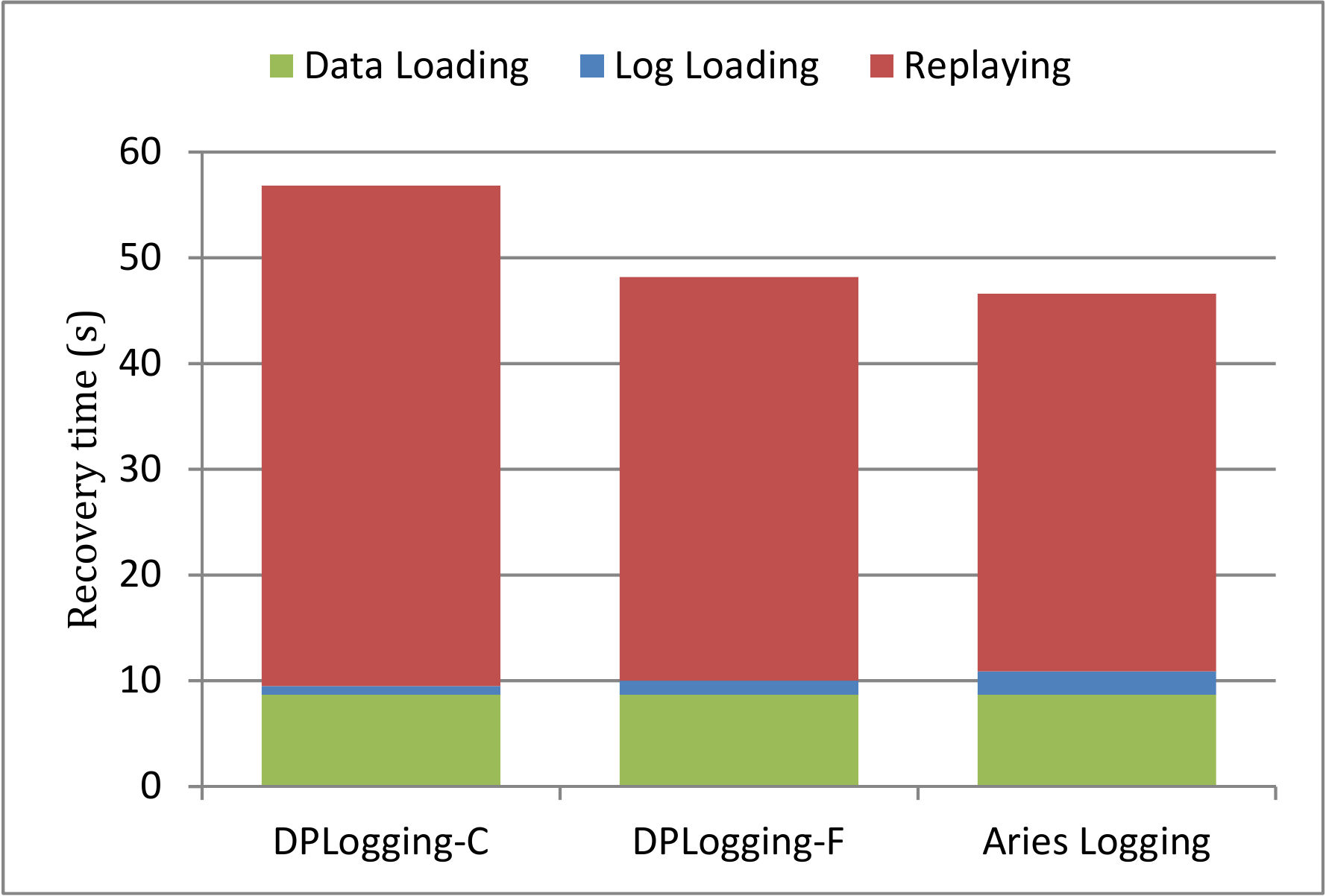}}\quad
	\subfigure[Recovery with 8 threads]{
		\label{fig:ycsb_recovery_8} 
		\includegraphics[scale=0.23]{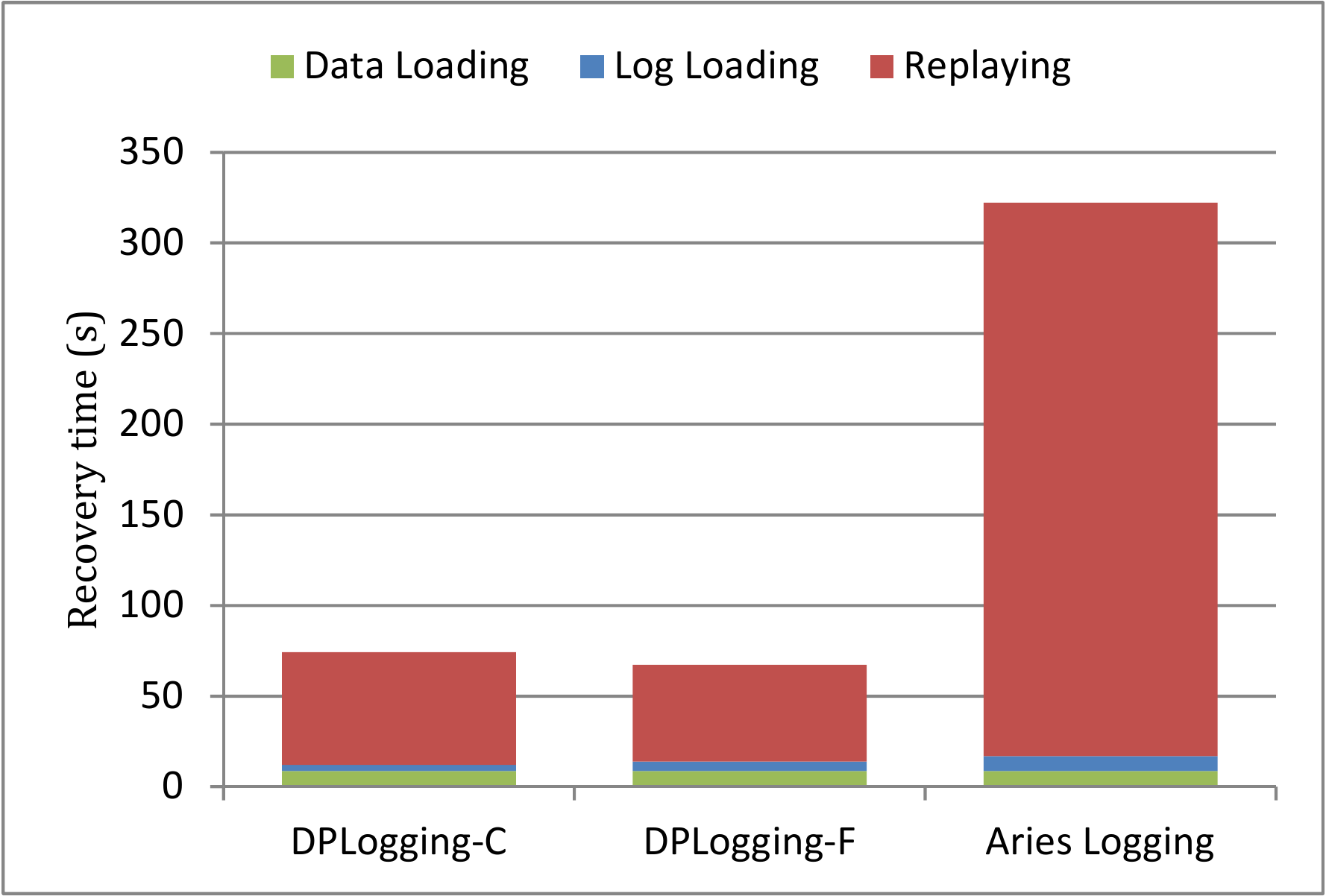}}
	\vspace{-3mm}
	\caption{Recovery time for single node on YCSB}
	\label{fig:ycsb_recovery_single_node}
\end{figure}

\begin{figure}[t]
	\subfigure[Recovery with 1 thread]{
		\label{fig:tpcc_recovery_1} 
		\includegraphics[scale=0.23]{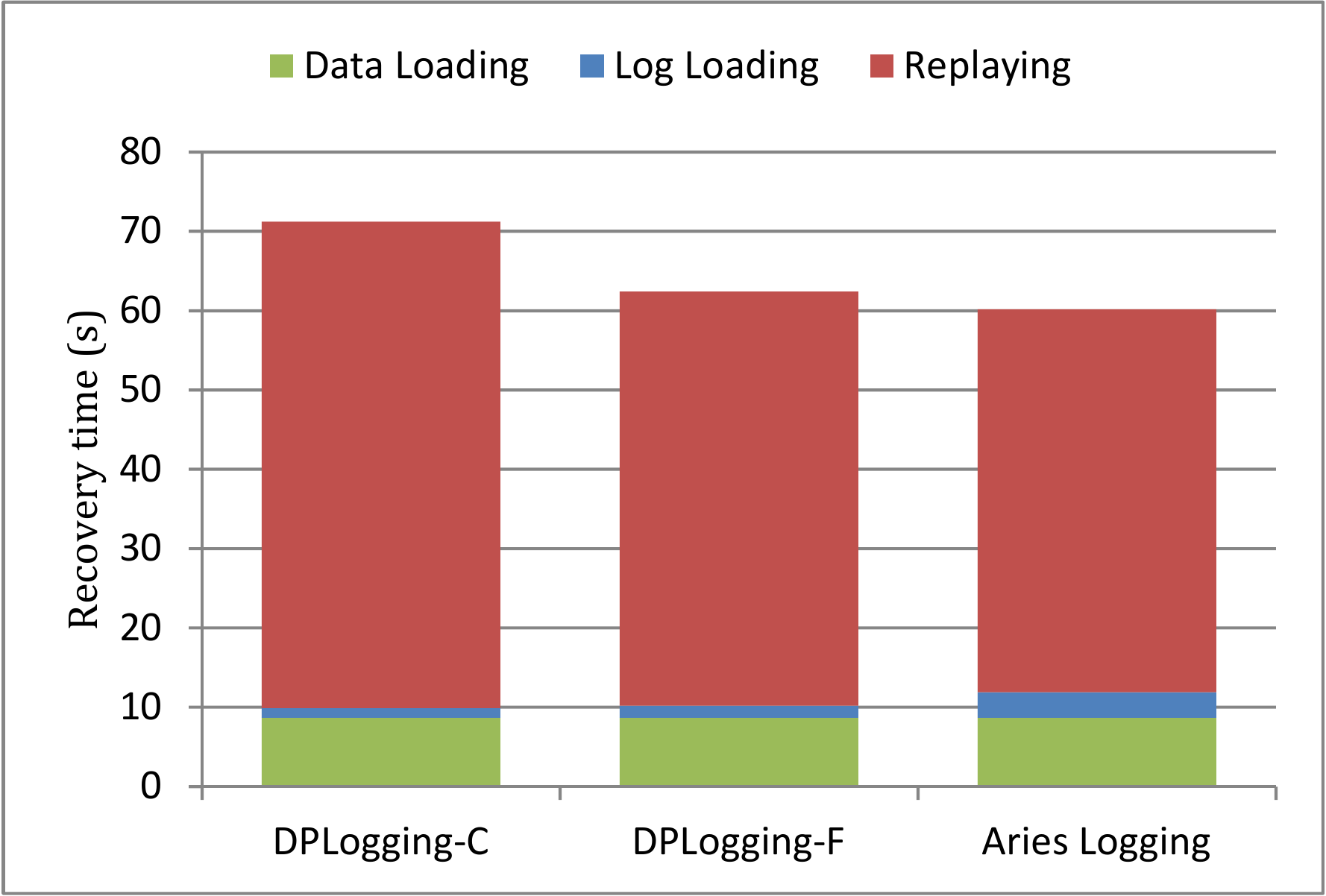}}\quad
	\subfigure[Recovery with 8 threads]{
		\label{fig:tpcc_recovery_8} 
		\includegraphics[scale=0.23]{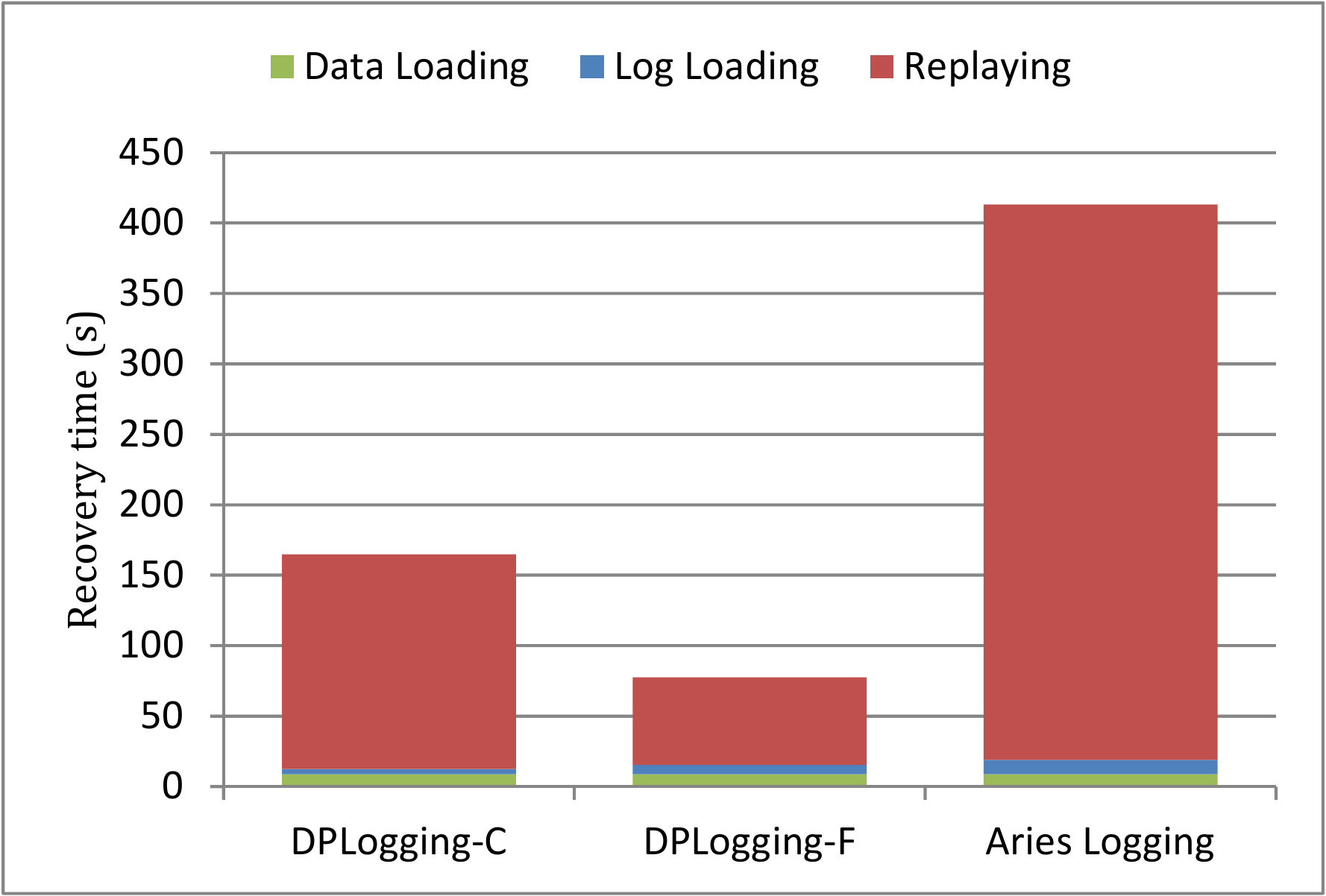}}
	\vspace{-3mm}
	\caption{Recovery time for single node on TPC-C}
	\label{fig:tpcc_recovery_single_node}
	\vspace{-2mm}
\end{figure}

\smallskip
\noindent\textbf{Recovery evaluation:}
We now evaluate the recovery performance with different logging approaches.
We simulate the failure by killing the daemon process of one node after the system runs for 60 seconds, and 
then we measure the time span for recovering the failed node.
Before the failed node starts to replay log records,
it must first load the latest database snapshot into memory.
Thus, the recovery time mainly consists of three parts:
data loading, log loading and replaying.
As shown in Figure~\ref{fig:ycsb_recovery_single_node} and Figure~\ref{fig:tpcc_recovery_single_node}, the time for data loading is almost the same for all logging approaches.
Compared to DPLogging-C and DPLogging-F,
Aries Logging spends more time to load the logs into memory,
since the log size of Aries Logging is much larger.

When only one worker thread is enabled, 
Aries Logging performs slightly better than DPLogging-C and DPLogging-F.
With only one worker thread available, all the three approaches replay their log records sequentially.
However, both DPLogging-C and DPLogging-F need to re-execute functions instead of updating the database directly. 
As Aries Logging saves data images before and after each update, 
read operations and transaction logics are not required to be redone during the recovery.
Thus, the recovery time with Aries Logging is smaller than DPLogging-F and DPLogging-C. 
When $8$ worker threads are enabled,
the recovery performance of Aries Logging drops significantly.
In the runtime, each worker thread maintains a log buffer to avoid contention.
However, in the recovery phase, all the log records in these log files have to replayed one by one,
since no dependency information can be used.   
Both DPLogging-C and DPLogging-F can achieve higher parallelism during the recovery by resolving
the dependency relations among log records.
DPLogging-F achieves the best recovery performance which is almost 5X faster than that of Aries Logging.
As shown in Figure~\ref{fig:ycsb_recovery_8},
DPLogging-C achieves a comparative performance to DPLogging-F on YCSB workload.
However, as shown in Figure~\ref{fig:tpcc_recovery_8}, 
DPLogging-F is 2X faster than that of DPLogging-C on TPC-C workload.
This is because each log record in DPLogging-C represents a transaction
and most transactions update the data record in the warehouse table, 
restricting the re-execution parallelism as a result.

\begin{figure}[t]
	\centering
	\subfigure[YCSB workload]{
		\label{fig:ycsb_recovery_running} 
		\includegraphics[scale=0.32]{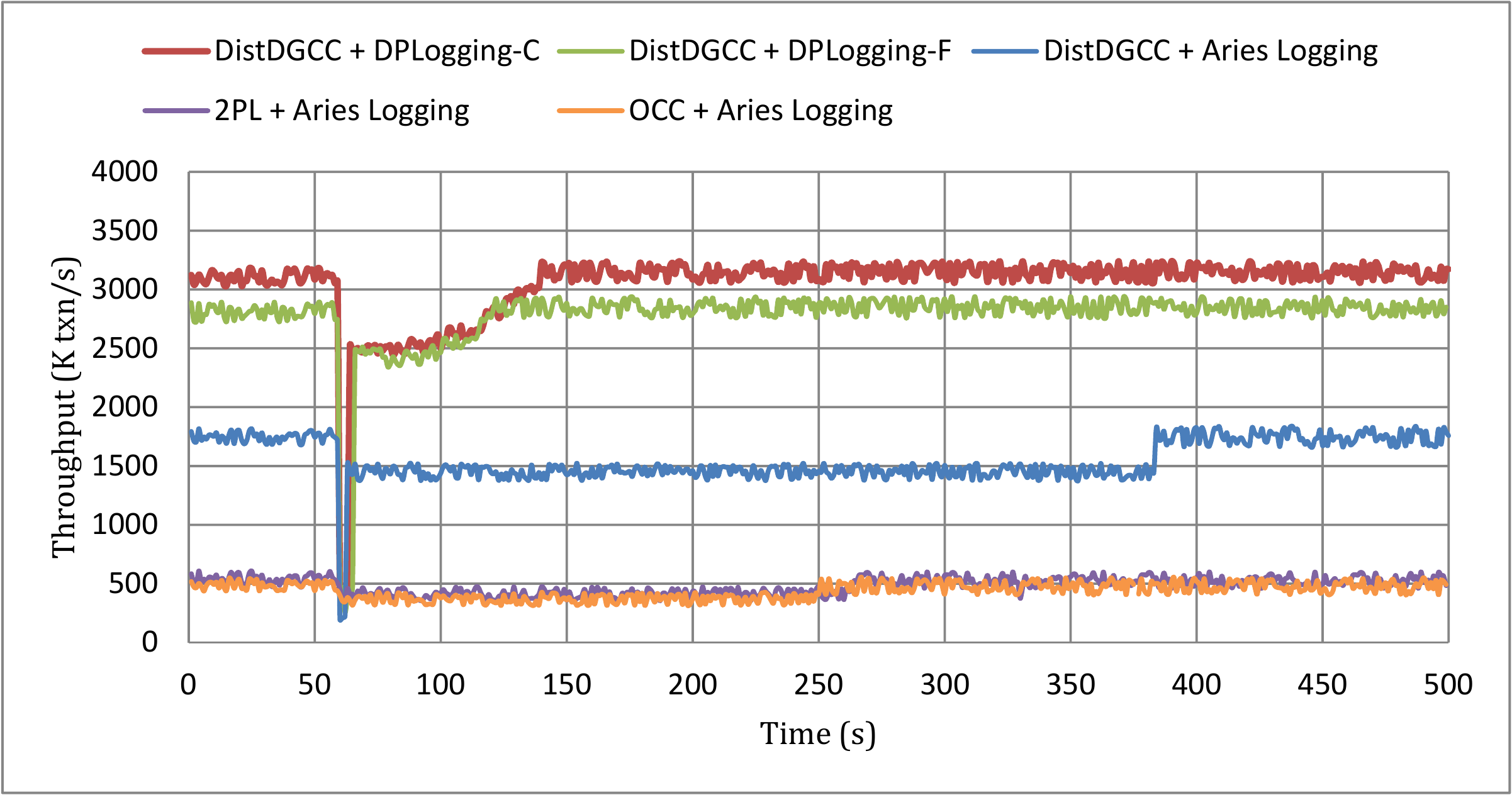}}
	\subfigure[TPC-C workload]{
		\label{fig:tpcc_recovery_running} 
		\includegraphics[scale=0.32]{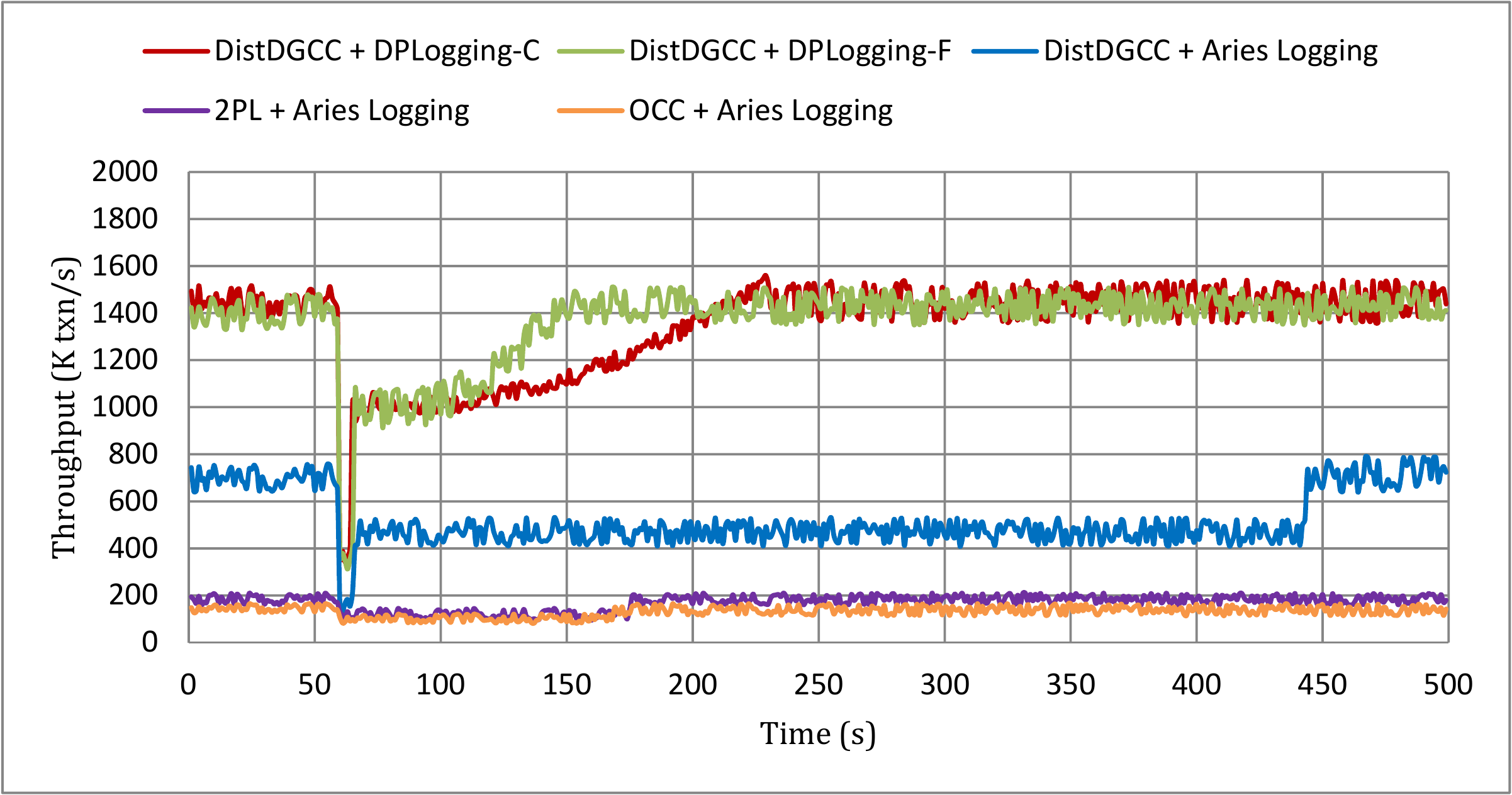}}
	\vspace{-2mm}
	\caption{Throughput during the recovery}
	\label{fig:recovery_running}
	\vspace{-3mm}
\end{figure}

\subsection{Overall Performance Evaluation}
In this section, we evaluate the overall performance of our proposed \distdgcc\ and dependency logging.
We run both YCSB and TPC-C workloads with 10\% distributed transactions for 500 seconds on the 8-node cluster.
After the systems run for 60 seconds, 
we kill the daemon process on a node randomly and invoke the recovery process.

Figure~\ref{fig:ycsb_recovery_running} and Figure~\ref{fig:tpcc_recovery_running} summarize the throughputs of the whole cluster.
When a failure occurs, system with Aries Logging cannot process transactions that access data on the failed node, and therefore abort them.
After the failed node is fully recovered, the throughput of the system returns to a normal level.
Compared to 2PL with Aries Logging and OCC with Aries Logging, 
we observe that \distdgcc\ with Aries Logging takes more time to complete the recovery process.
This is because \distdgcc\ is more robust to distributed transactions and has a higher runtime throughput.
Running for the same amount of time (60 seconds in the experiment), 
\distdgcc\ commits more transactions before the failure.
Thus, it requires more time to replay the committed log records.
For systems with dependency logging (both DPLogging-C and DPLogging-F),
the failed node can be recovered according to the dependency graphs 
and process new incoming transactions at the same time.
Thus, the throughput gradually increases as the failed node is being recovered.
As discussed above, transactions in TPC-C workload contend on a small set of data records.
Hence, system with DPLogging-C takes more time to recover than system with DPLogging-F.

As shown in Figure~\ref{fig:recovery_running},
\distdgcc\ is efficient in distributed environment and dependency logging supports fast recovery with marginal runtime overhead.
As a consequence, 
\distdgcc\ with dependency logging exhibits superior overall performance compared to the state-of-the-art techniques.


\section{Related Work}
\label{sec:related_work}
	
\noindent\textbf{Concurrency Control Protocols.}
High efficient concurrency control protocol that ensures correct execution of concurrent transactions is vital for in-memory database systems.
Two-phase locking (2PL)~\cite{cacm76:Eswaran} and Optimistic Concurrency Control (OCC)~\cite{tods81:Kung} are most widely adopted.
As a pessimistic protocol, 2PL needs to acquire the lock before accessing a tuple and release it after transaction commits or aborts.
With 2PL, conflict operations are resolved in advance and are executed in sequence.
On the contrary, OCC assumes that conflicts are rare and does not check the conflicts during the transaction execution.
Each transaction maintains read and write sets and conducts a conflict validation.
Transaction commits only when the validation phase is passed, otherwise it restarts or directly aborts.
With the advancement of new hardware techniques, many research efforts have been devoted to improve the efficiency of concurrency control protocols.

In Multi-Version Concurrency Control (MVCC)~\cite{bernstein1987,sigmod1992:mohan}, read operation does not block write operations. Hekaton~\cite{sigmod13:Diaconu,vldb2011:larson} makes use of MVCC together with a lock-free hash table to improve it performance. Hyper~\cite{icde11:Kemper,sigmod2015:neumann} and BOHM~\cite{vldb15:Faleiro} extends MVCC to enforce serializability and avoid shared memory writes for read tracking, respectively.  
For lock-based concurrency control protocols, the lock manager is typically very complex and incurs performance bottleneck.
Light-weight Intent Lock (LIL)~\cite{vldb2012:kimura} introduces light-weight counter in a global lock table to ease the management. 
Very Lightweight Lock (VLL)~\cite{vldb2012:ren} maintains a lock state with each tuple and removes centralized lock manager.
However, LIL has to block transaction that cannot obtain all the locks and the performance of VLL is seriously affected by workloads that cannot be well partitioned.
Silo~\cite{sosp13:Tu} optimizes OCC by adopting a batched timestamp allocation.
\cite{sigmod2016:wu,sigmod2016:yu,sigmod2016:kim,sigmod2015:kimura,vldb2016:yuan} show that OCC suffers for high contention workloads.
They identify the bottlenecks and propose new approaches to improve its performance.
However, all these optimizations mainly focus on multi-core systems rather than distributed systems.
H-Store~\cite{vldb08:Kallman} and Hyper~\cite{icde11:Kemper} adopt single-threaded model on partitioned databases to eliminate the overhead caused by concurrency control.
However, their performance may suffer for workloads with more cross-partition transactions.

\smallskip

\noindent\textbf{Fault-Tolerant Schemes.}
ARIES~\cite{tods92:Mohan} is the most widely used logging approach in traditional database systems.
By maintaining data in memory, new recovery techniques~\cite{vldb1994:jagadish, vldb1993:jagadish, fofo1989:lehman, osdi2014:zheng} have been put forward, most of which inherits the idea from ARIES.
Logical logging techniques~\cite{icde14:Malviya,vldb2011:lomet} are proposed recently that aim to reduce the log size.
While they improve the runtime performance by reducing the number of disk I/Os to an extent, they usually incur expensive cost for recovery, especially in distributed environment~\cite{sigmod2016:yao}.
Many optimizations are also proposed to increase the efficiency of logging and recovery.
\cite{icde1993:li} makes use of shadow pages to reduce the log size during the runtime.
\cite{vldb2010:johnson} reduces the lock contention on the log buffers and the effects of context switching to improve the logging performance.

Recently, many research efforts have also been devoted to combine logging techniques with Non-Volatile Memory (NVM).
\cite{vldb2013:pelley} attempts to reduce the number of writes on NVM and \cite{vldb2014:wang} introduces a distributed logging protocol on NVM by making use of group commit approaches.
Write-Behind Logging~\cite{vldb2016:arulraj} flushes changes of databases before flushing the logs by tracking where the databases are changed instead of how they are changed.
Since our design only considers on a general commodity machines, we do not apply these techniques.
	
Replication-based techniques~\cite{sosp2011:ongaro,vldb2005:stonebraker} provide fast recovery and high availability.
However, these techniques incur expensive coordination cost during the runtime to achieve strong consistency.
Moreover, given a limited amount of memory, it is expensive to maintain replicas, especially for very large databases.
In most cases, replication-based approaches do not work for the whole cluster failure.
Further, logging techniques are usually are orthogonal to replication-based techniques.

\section{Conclusion}
\label{sec:conclusion}

In this paper, we designed transaction management and logging within the same framework.
We proposed a distributed graph based concurrency control protocol that reduces the cost for distributed transactions by aggregating network messages for a batch of transactions.
Subsequently, we also proposed
 a new dependency logging technique, and associated fine-grained dependency logging and coarse-grained dependency logging approaches.
By tracking the dependency relations among the transactions, dependency logging enables parallel recovery that helps to speed up the recovery and reduce the system downtime.
Extensive experiments on both YCSB and TPC-C workloads confirm that our system with dependency based partitioning distributed transaction management and logging exhibits superiority in both runtime throughout and recovery performance.

\section{Acknowledgments}
This work was in part supported by the National Research
Foundation, Prime Minister's Office, Singapore under its
Competitive Research Programme and Tencent Grant.

{
\scriptsize
\bibliographystyle{abbrv}
\qquad
\bibliography{vldb_sample}
}
	
\end{document}